\documentclass[preprint,12pt,authoryear]{elsarticle}
\pdfoutput=1 

\RequirePackage[OT1]{fontenc}
\RequirePackage{amsthm,amsmath}
\usepackage[font=scriptsize]{caption}
\usepackage{adjustbox}
\usepackage{enumerate}
\usepackage{amsmath}
\usepackage{amsfonts}
\usepackage{amssymb}
\usepackage{amsthm}
\usepackage[makeroom]{cancel}
\usepackage{bbold}
\usepackage{float}
\usepackage[textsize=tiny]{todonotes}
\usepackage{algorithm}
\usepackage{algpseudocode}

\usepackage{url} 
\usepackage[colorlinks=true,
            linkcolor=blue,
            urlcolor=blue,
            citecolor=blue]{hyperref}
\usepackage{etoolbox}
\usepackage{comment}
\usepackage{array}
\usepackage{bm}
\newcolumntype{H}{>{\setbox0=\hbox\bgroup}c<{\egroup}@{}}
\newcommand{\B}[1]{\mathbf{#1}}
\newcommand{\BB}[1]{\boldsymbol{#1}}
\usepackage{xr}
\AtBeginEnvironment{tabular}{\small}

\newcommand{\Bgamma}{\boldsymbol{\mathsf{\gamma}}}
\newcommand{\By}{\boldsymbol{y}}
\newcommand{\Bchi}{\boldsymbol{\chi}}

\newtheorem{theorem}{Theorem}

\usepackage{pdfcomment}
\pdfcommentsetup{draft} 
\pdfcommentsetup{color={1.0 1.0 0.0}, open=true}

\begin{document}

  \title{Mode jumping MCMC for Bayesian variable selection in GLMM}
  \author{Aliaksandr Hubin \footnote{The corresponding author, Aliaksandr Hubin, is a PhD candidate at the University of Oslo, 0851 Moltke Moes vei 35 Oslo, Norway. Email: aliaksah@math.uio.no, tel: +4745171361.\\The paper has supplementary materials consisting of:\\ \textbf{R package:} \textit{R} package \textit{EMJMCMC} to perform the efficient mode jumping MCMC described in the paper. (EMJMCMC\_ 1.2.tar.gz; GNU zipped tar file).\\\textbf{Data and code:} Data (simulated and real) and \textit{R} code for MJMCMC algorithm, post-processing and creating figures wrapped together into a reference based EMJMCMC class. (code-and-data.zip; zip file containing the data, code and a read-me file (readme.pdf))\\
\textbf{Details and Pseudo code:} Proofs of the ergodicity of MJMCMC procedure, pseudo codes for MJMCMC and local combinatorial optimizers, parallelization strategies and some supplementary tables for the experiments. (appendix.pdf)
}\hspace{.2cm}\\
    Department of Mathematics, University of Oslo\\
    and \\
    Geir Storvik \\
    Department of Mathematics, University of Oslo}

\begin{frontmatter}
\begin{abstract}
Generalized linear mixed models (GLMM) are used for inference and prediction in a wide range of different applications providing a powerful scientific tool. An increasing number of sources of data are becoming available, introducing a variety of candidate explanatory variables for these models. Selection of an optimal combination of variables is thus becoming crucial. In a Bayesian setting, the posterior distribution of the models, based on the observed data, can be viewed as a  relevant measure for the model evidence.
The number of possible models increases exponentially in the number of candidate variables. Moreover, the space of models has numerous local extrema in terms of posterior model probabilities. To resolve these issues a novel MCMC algorithm for the search through the model space via efficient mode jumping for GLMMs is introduced. The algorithm is based on that marginal likelihoods can be efficiently calculated within each model. It is recommended that either exact expressions or precise approximations of marginal likelihoods are applied. The suggested algorithm is applied to simulated data, the famous U.S. crime data, protein activity data and epigenetic data and is compared to several existing approaches.
\end{abstract}


\end{frontmatter}

\section{Introduction}\label{section1}

In this paper we study variable selection in generalized linear mixed models (GLMM) addressed in a Bayesian setting. 
Being one of the most powerful modeling tools in modern statistical science \citep{Ariti29122014} these models  have proven to be efficient in numerous applications including simple banking scoring problems \citep{Grossi2006}, insurance claims modeling \citep{David2015147}, studies on the course of illness in schizophrenia, linking diet with heart diseases \citep{Skrondal2003}, analyzing sophisticated astrophysical data \citep{deSouza:2014loa}, and inferring on genomics data \citep{Lobréaux201569}. In many of these applications, the number of  candidate explanatory variables (covariates) is large,  making  variable selection  a difficult problem, both conceptually and numerically. In this paper we will focus on efficient Markov chain Monte Carlo (MCMC) algorithms for such variable selection problems. Our focus will be on posterior model probabilities  although other model selection criteria can also easily be adopted within the algorithm.

Algorithms for variable selection in the Bayesian settings have been previously addressed, but primarily in the combined space of models \emph{and} parameters.  \citet{Goerge2007Search} describe and compare various hierarchical mixture prior formulations for Bayesian variable selection in normal linear regression models. They outline computational methods including Gray Code sequencing and standard MCMC for posterior evaluation and exploration of the space of models. They also comment on the infeasibility of exhaustive exploration of the space of models for moderately large problems as well as the inability of standard MCMC techniques to escape from local optima efficiently. \citet{AlAwadhi2004189} consider using several MCMC steps within a new model to obtain good proposals within the combined parameter and model domain while~\citet{YehYang2012} propose local annealing approaches. \citet{WICS1352} also addresses MCMC algorithms to estimate the posterior distribution over models. She observes that estimates of posterior probabilities of individual models based on MCMC output are often not reliable because the number of MCMC samples is typically considerably smaller than the size of the model space. As a consequence she considers the median probability model of~\citet{barbieri2004optimal} instead and shows that this algorithm can, under some conditions, outperform standard  MCMC. 
Yet another approach for Bayesian model selection is addressed by \citet{Bottolo15022011}, who propose the moves of MCMC between local optima through a permutation based genetic algorithm that has a pool of solutions in a current generation suggested by parallel tempered chains.
A similar idea is considered by \citet{Frommlet2012}.
Multiple try MCMC methods with local optimization have been described by \citet{Liu2000eMTMCMC}. 
\citet{RSSB:RSSB12095} address the case when there is by far more explanatory variables than observations. They suggest a split and merge Bayesian model selection algorithm that first splits the set of covariates into a number of subsets, then finds relevant variables from these subsets and in the second stage merges these relevant variables and performs a new selection from the merged set. This algorithm in general cannot guarantee convergence to a global optimum or find the true posterior distribution of the models, however under some strict regularity conditions it does so asymptotically.

For an increasing number of  model classes,  marginal likelihoods for specific models  can be efficiently  calculated, either exactly or approximately. This makes the exploration of models much easier.  \cite{bove2011bayesian} consider an MCMC algorithm within the model space, but only allow local moves. This might be a severe limitation in cases where multiple sparsely located modes are present in the model space.
\citet{bivand2014approximate} combine approximations of marginal likelihood with Bayesian model averaging within spatial models. 
\citet{Clyde:Ghosh:Littman:2010} suggest a Bayesian adaptive sampling (BAS) algorithm as an alternative to MCMC allowing for perfect sampling without replacement.  

In the general MCMC literature, various algorithms for exploration of model spaces with multiple sparse modes have been suggested. These approaches can be divided into two groups: methods based on exploration of the tempered target distributions (allowing to flatten or increase multimodality for different temperatures) and methods based on utilization of local gradients. The first group of algorithms was initialized with the parallel tempering approach \citep{geyer1991markov}, which further had numerous modifications \citep{liang2010double, miasojedow2013adaptive, salakhutdinov2009learning}. One of the most prominent extensions is the equi-energy sampling approach \citep{kou2006discussion}, which utilizes the physical duality between temperature and energy. This approach targets directly the former to flatten or tighten the parameter spaces. Another extension is the multi domain sampling approach \citep{zhou2011multi}, which first uses the target distribution tempering idea to find the set of local modes and then uses local MCMC to explore the regions around them for further global inference. The second group of algorithms uses auxiliary variables combined with gradients of the extended distribution to explore the state space accurately~\citep[][and many others]{neal2011mcmc, chen2014stochastic, sengupta2016gradient}. Both groups of algorithms are mainly developed for exploration of continuous parameter spaces. All of these algorithms can in principle be adapted to discrete space problems. 
The approach in this article will be to adapt the mode jumping MCMC idea of \citet{Tjelmeland99modejumping} to the variable selection problem, utilizing 
the existence of marginal likelihoods for models of interest.

Different approaches can be applied for calculation of marginal likelihoods. For linear models with conjugate priors, analytic expressions are available~\citep{Clyde:Ghosh:Littman:2010}. In more general settings, MCMC algorithms combined with e.g. Chib's method~\citep{chib1995marginal} can be applied, giving however computationally expensive procedures.  See also~\citet{Friel2012} for alternative MCMC based methods. For Gaussian latent variables, the computational task can be efficiently solved through the integrated nested Laplace approximation (INLA) approach \citep{rue2009eINLA}. \citet{HubinStorvikINLA} compare INLA with MCMC based methods, showing that INLA based approximations are extremely accurate and require much less computational effort than the MCMC approaches for within-model calculations.

In this paper we introduce a novel MCMC algorithm for search through the model space, the mode jumping MCMC (MJMCMC). 
The focus will be on Gaussian latent variable models, for which efficient approximations to marginal likelihoods are available. The algorithm is based on the idea of mode jumping within MCMC - resulting in an MCMC algorithm which manages to efficiently explore the model space by means of mode jumping, applicable through large jumps combined with
local optimization.
Mode jumping MCMC methods within a continuous space setting were first suggested by~\citet{Tjelmeland99modejumping}. 
We modify the algorithm to the discrete space of possible models, requiring both new ways of making large jumps and of performing local optimization.  
We include mixtures of proposal distributions and parallelization to further improve the performance of the algorithm.  A valid acceptance probability within the Metropolis-Hastings setting is constructed based on the use of backward kernels.

\section{The generalized linear mixed model}\label{section2}
We consider the following generalized linear mixed model: 
\begin{align} \label{themodeleq}
  Y_i|\mu_i \sim&  \mathfrak{f}(y|\mu_i),
  \quad \mu_i = g^{-1}\left(\eta_i\right), \\
   \eta_i =&  \beta_0 + \sum_{j=1}^{p} \gamma_j\beta_{j}x_{ij} + \delta_i
   \intertext{and}
 \boldsymbol{\delta} =& (\delta_1,...,\delta_n) \sim N_n\left(\boldsymbol{0},\boldsymbol{\Sigma}_b\right).\label{themodeleqend}
\end{align}
Here $Y_i$ is the response variable while $x_{ij}, j =1,...,p$ are the covariates. We assume $\mathfrak{f}(y|\mu)$ is a density/distribution from the exponential family with corresponding link function $g(\cdot)$.  The latent indicators $\gamma_j\in\{0,1\}, j = 1,...,p$ define if covariate $x_{ij}$ is  to be included into the model ($\gamma_j = 1$) or not ($\gamma_j = 0$) while
$\beta_j \in \mathbb{R}, j=0,...,p$ are the corresponding regression coefficients.  We are also addressing the unexplained variability of the responses and the correlation structure between them through random effects $\delta_i$ with a specified parametric covariance matrix structure defined through $\boldsymbol{\Sigma}_b = \boldsymbol{\Sigma}_b\left(\boldsymbol{\psi}\right) \in \mathbb{R}^{n\times n}$, where $\boldsymbol{\psi}$ are parameters describing the correlation structure.

In order to put the model into a Bayesian framework, we assume
\begin{align}
\gamma_j|q \sim& \text{Binom}(1,q),\quad j=1,...,p\label{glmgammaprior}
\intertext{and}
q \sim& \text{Beta}(a_q,b_q),\label{glmgammahyperprior}
\end{align}
where $q$ is the prior probability of including a covariate into the model. For $(\boldsymbol{\beta},\boldsymbol{\psi})$ different priors are possible, see the applications in section~\ref{section4}.

Let $\Bgamma = (\gamma_1,...\gamma_p)$, which uniquely defines a specific model. Assuming the constant term $\beta_0$ is always included, there are $L = 2^{p}$ different models to consider. We want to find a set of the best models with respect to posterior model probabilities $p(\Bgamma|\By)$,  where $\By=(y_1,...,y_n)$. 
We assume that marginal likelihoods $p(\By|\Bgamma)$
are available for a given $\Bgamma$, and then use MCMC to explore 
$p(\Bgamma|\By)$. By Bayes formula 
\begin{equation}\label{PMP}
p(\Bgamma|\By) =  \frac{{p(\By|\Bgamma)p(\Bgamma)}}{\sum_{\Bgamma' \in\Omega}{p(\By| \Bgamma')p(\Bgamma')}}.
\end{equation}
In order to calculate $p(\Bgamma|\By)$ we have to iterate through the whole model space $\Omega$, which becomes computationally infeasible for large $p$. 
The ordinary MCMC based estimate is based on a number of MCMC samples $\Bgamma^{(i)},i= 1,...,W$:
\begin{equation}\label{map2}
\widetilde{p}(\Bgamma|\By)=\frac{\sum_{i=1}^{W}{\mathbb{1}(\Bgamma^{(i)} = \Bgamma)}}{W} \xrightarrow[W\rightarrow\infty]{d} p(\Bgamma|\By),
\end{equation}
where $\mathbb{1}(\cdot)$ is  the indicator function. 
An alternative,  named the renormalized model (RM) estimates by \citet{Clyde:Ghosh:Littman:2010}, is
\begin{equation}
\widehat{p}(\Bgamma|\By) =  \frac{{p(\By|\Bgamma)p(\Bgamma)}}{\sum_{\Bgamma' \in \mathbb{V}}{p(\By| \Bgamma')p(\Bgamma')}}\mathbb{1}(\Bgamma \in \mathbb{V}),\label{approxpost}
\end{equation}
where now $\mathbb{V}$ is the set of visited models during the
MCMC run. Although both~\eqref{approxpost} and~\eqref{map2} 
are asymptotically consistent,  \eqref{approxpost} 
will often be the preferable estimator since convergences of the MCMC based approximation~\eqref{map2} 
is much slower, see~\citet{Clyde:Ghosh:Littman:2010}.

We aim at approximating $p(\Bgamma|\By)$ by means of searching for some subspace $\mathbb{V}$ of $\Omega$ making the approximation~\eqref{approxpost} as precise as possible.
Models with high values of $p(\By|\Bgamma)$ are important to be addressed.  This means that modes and near modal values of marginal likelihoods are particularly important for construction of reasonable $ \mathbb{V} \subset  \Omega$ and missing them can dramatically influence our estimates. Note that these are aspects just as important if̃ the standard MCMC estimate~\eqref{map2} is to be used. A main difference is that while for using~\eqref{map2} the number of times a specific model is visited is important, for~\eqref{approxpost} it is enough that a model is visited at least once.
In this context the denominator of \eqref{approxpost}, which we would like to be as high as possible, becomes an extremely relevant measure for the quality of the search in terms of being able to capture whether the algorithm visits all of the modes, whilst the size of $\mathbb{V}$ should be low in order to save computational time.

The posterior marginal inclusion probability $p(\gamma_j=1|\By)$ can be approximated by
\begin{equation}\label{margininuspost}
\widehat{p}(\gamma_j=1|\By) = \sum_{\Bgamma' \in \mathbb{V}}{\mathbb{1}(\gamma_j'=1){\widehat p}(\Bgamma'|\By)},
\end{equation}
giving a measure for assessing importance of the covariates. Other parameters can be estimated similarly.

Algorithms for estimating $\mathbb{V}\,$ are described in section~\ref{section3}.
In practice $p(\By|\Bgamma)$  may not be available analytically. We then rely on some precise approximations $\widehat{p}(\By|\Bgamma)$. Such approximations introduce additional errors in~\eqref{approxpost} and~\eqref{margininuspost}, but we assume them to be small enough to be ignored. This is further discussed in section~\ref{sec:marg.dens}.

\section{Mode jumping Markov chain Monte Carlo }\label{section3}
MCMC algorithms \citep{robert2004bayes} have been extremely popular for the exploration of model spaces for model selection, being capable of providing samples from the posterior distribution of the models. 
In our setting, the most important aspect becomes building a method to explore the model space in a way to efficiently switch between potentially sparsely located modes, whilst avoiding visiting models with a low $p(\By|\Bgamma)$ too often.

\subsection{Standard Metropolis-Hastings}

Metropolis-Hastings algorithms \citep{robert2004bayes} are a class of MCMC methods for drawing from a complicated target distribution living on some space $\Omega$, which in our setting will be $\pi(\Bgamma)=p(\Bgamma|\By)$.
 Given some proposal distribution $q(\Bgamma^*|\Bgamma)$,
the Metropolis-Hastings algorithm accepts the proposed $\Bgamma^*$  with probability
\begin{equation}
r_{mh}(\Bgamma,\Bgamma^*)=\min\left\lbrace1,\frac{\pi(\Bgamma^*)q(\Bgamma|\Bgamma^*)}{\pi(\Bgamma)q(\Bgamma^*|\Bgamma)}\right\rbrace,\label{balance}
\end{equation}
and otherwise remains in the old state $\Bgamma$.
This will generate a Markov chain which, given the chain is irreducible and aperiodic, will have $\pi$ as stationary distribution. Theoretical results related to convergence of MCMC based estimates can be found in e.g.~\citet{tierney1996introduction}. Note that the discrete finite space of models make these results easily applicable in our case.

Given that the $\gamma_j$'s are binary, changes correspond to swaps between the values 0 and 1.
One can address various options for generating proposals. A simple proposal is
to first select the number of components to change, e.g. $S\sim \text{Unif}\{\zeta,...,\eta\}$, followed by a sample of size $S$ without replacement from $\{1,...,p\}$.
This implies that in \eqref{balance} the proposal probability for switching from $\Bgamma$ to $\Bgamma^*$ becomes symmetric, simplifying calculation of the acceptance probability.
Other possibilities for proposals are summarized in Table~\ref{proposals}, allowing, among others, different probabilities of swapping for the different components. Such probabilities can for instance be associated with marginal inclusion probabilities from a preliminary MCMC run.

\begin{table}[t]
\begin{adjustbox}{center}
\begin{tabular}{cll}
\hline\hline
\textbf{Type}&Proposal
&\textbf{Label}\\\hline
\\[1pt]
1&$\frac{\prod_{j\in \{j_1,...,j_S\}}\rho_{j}}{\binom {p} {S}(\eta-\zeta +1)}$&\textit{Random change with random size of the neighborhood}\\
2&$\frac{\prod_{j\in \{j_1,...,i_S\}}\rho_{j}}{\binom {p} {S}}$&\textit{Random change with fixed size of the neighborhood}\\
3&$\frac{1}{\binom {p} {S}(\eta-\zeta +1)}$&\textit{Swap with random size of the neighborhood}\\
4&${\binom {p} {S}}^{-1}$&\textit{Swap with fixed size of the neighborhood}\\
5&$\frac{1-\mathbb{1}\left(\sum_{j}^{p}\gamma_j=p\right)}{p-\sum_{j}^{p}\gamma_j + \mathbb{1}\left(\sum_{j}^{p}\gamma_j=p\right)}$&\textit{Uniform addition of a covariate}\\
6&$\frac{1-\mathbb{1}\left(\sum_{j}^{p}\gamma_j=0\right)}{\sum_{j}^{p}\gamma_j + \mathbb{1}\left(\sum_{j}^{p}\gamma_j=0\right)}$&\textit{Uniform deletion of a covariate}\\\hline
\end{tabular}\end{adjustbox}
\\[1pt]
\caption{Types of proposals suggested for moves between models during an MCMC procedure. Here $S$ is either a deterministic or random ($S\sim{Unif}\{\zeta,...,\eta\}$) size of the neighborhood; $\rho_{j}$ is the probability of a change on variable $\gamma_j$.}\label{proposals}
\end{table} 

\subsection{MJMCMC - the mode jumping MCMC}\label{mjmcmcsimp}

The main problem  with  the standard Metropolis-Hastings algorithms is the trade-off between  possibilities of large jumps (by which we understand proposals with a large neighborhood) and  high acceptance  probabilities. Large jumps will typically result in  proposals with low probabilities.
In a continuous setting, \citet{Tjelmeland99modejumping} solved this  by  introducing  local optimization  after  large jumps, which results in proposals with higher acceptance probabilities.  We adapt this approach to the  discrete model selection setting by the following  algorithm:
\begin{algorithm}[H]
\caption{Mode jumping MCMC}\label{MJMCMCalg0}
\begin{algorithmic}[1]
\State  Generate a large jump  $\Bchi_0^*$ according to a  proposal distribution  $q_l(\Bchi_0^*|\Bgamma)$.
\item Perform a local  optimization, defined through $\Bchi_k^*\sim q_o(\Bchi_k^*|\Bchi_0^*)$.
\State Perform a small randomization  to generate the proposal $\Bgamma^*\sim q_r(\Bgamma^*|\Bchi_k^*)$.
\item  Generate backwards auxiliary variables $\Bchi_0\sim q_l(\Bchi_0|\Bgamma^*)$,  $\Bchi_k\sim q_o(\Bchi_k|\Bchi_0)$.
\State Put
\[\Bgamma'=
\begin{cases}\Bgamma^*&\text{with probability $r_{mh}(\Bgamma,\Bgamma^*;\Bchi_k,\Bchi_k^*)$;}\\
\Bgamma&\text{otherwise,}
\end{cases}
\]
where
\begin{equation}
r_{mh}^*(\Bgamma,\Bgamma^*;\Bchi_k,\Bchi_k^*) = \min\left\{1,\frac{\pi(\Bgamma^*)q_r(\Bgamma|\Bchi_{k})}{\pi(\Bgamma)q_r(\Bgamma^*|\Bchi_k^*)}\right\}\label{locmcmcgen00}.
\end{equation}
\end{algorithmic}
\end{algorithm}
Here a large jump corresponds to changing a large number of $\gamma_j$'s while the local optimization will be some iterative procedure based on, at each iteration, changing a small number of components until a local mode is reached.

\begin{figure}[t]
    \centering
    \includegraphics[width=1.0\textwidth]{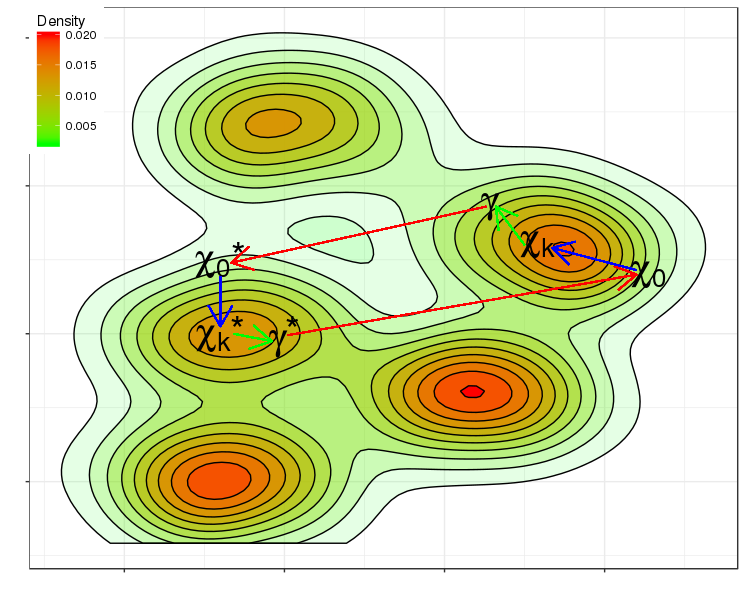}
    \caption{Graphical illustration of a MJMCMC step with a large jump followed by a locally optimized proposal. The red arrows correspond to the large jumps, the blue arrows correspond to local optimization, the green arrows correspond the randomization steps.}
    \label{fig:opt0}
\end{figure}

The procedure is illustrated in Figure~\ref{fig:opt0} where the backward sequence $\Bgamma^*\rightarrow\Bchi_0\rightarrow\Bchi_k\rightarrow\Bgamma$, needed for calculating the acceptance probability, is included. 
 For this algorithm, three proposals need to be specified; $q_l(\cdot|\cdot)$ specifying the first large jump,
$q_o(\cdot|\cdot)$ specifying the local optimizer, and $q_r(\cdot|\cdot)$ specifying the last randomization, all to be described in more details below.

$\pi$-invariance of the MJMCMC procedures is given by the following  theorem \citep[based on similar arguments as in][]{geirs2011mhf,Chopin2013}:
\begin{theorem}
Assume $\Bgamma\sim \pi(\cdot)$ and $\Bgamma'$ is generated according to   Algorithm~\ref{MJMCMCalg0}. 
Then  $\Bgamma'\sim \pi(\cdot)$.
\end{theorem}

\begin{proof}
Since $\Bgamma\sim \pi(\cdot)$ and $(\Bchi_0^*,\Bchi_k^*)\sim q_l(\Bchi_0^*|\Bgamma)q_o(\Bchi_k^*|\Bchi_0^*)$ we have that
\[
(\Bgamma,\Bchi_0^*,\Bchi_k^*)\sim \pi(\Bgamma)q_l(\Bchi_0^*|\Bgamma)q_o(\Bchi_k^*|\Bchi_0^*)
\equiv \bar{\pi}(\Bgamma,\Bchi_0^*,\Bchi_k^*).
\]
We may now consider $(\Bgamma^*,\Bchi_0,\Bchi_k)$ as a proposal in the extended space, generated according to the distribution $q_r(\Bgamma^*|\Bchi_k^*)q_l(\Bchi_0|\Bgamma^*)q_o(\Bchi_k|\Bchi_0)$.  An ordinary  Metropolis-Hastings iteration with respect to  $\bar{\pi}(\Bgamma,\Bchi_0^*,\Bchi_k^*)$ is then to accept $(\Bgamma^*,\Bchi_0,\Bchi_k)$ with probability
$r_{mh}^*=\min\{1,\alpha_{mh}^*\}$ where
\begin{align*}
\alpha_{mh}^*
=&\frac
  {\bar{\pi}(\Bgamma^*,\Bchi_0,\Bchi_k)q_r(\Bgamma|\Bchi_k)q_l(\Bchi_0^*|\Bgamma)q_o(\Bchi_k^*|\Bchi_0^*)}
  {\bar{\pi}(\Bgamma,\Bchi_0^*,\Bchi_k^*)q_r(\Bgamma^*|\Bchi_k^*)q_l(\Bchi_0|\Bgamma^*)q_o(\Bchi_k|\Bchi_0)}\\
  =&\frac
  {\pi(\Bgamma^*)q_l(\Bchi_0|\Bgamma^*)q_o(\Bchi_k|\Bchi_0)
  q_r(\Bgamma|\Bchi_k)q_l(\Bchi_0^*|\Bgamma)q_o(\Bchi_k^*|\Bchi_0^*)}
  {\pi(\Bgamma)q_l(\Bchi_0^*|\Bgamma)q_o(\Bchi_k^*|\Bchi_0^*)
  q_r(\Bgamma^*|\Bchi_k^*)q_l(\Bchi_0|\Bgamma^*)q_o(\Bchi_k|\Bchi_0)}
  =\frac
  {\pi(\Bgamma^*)q_r(\Bgamma|\Bchi_k)}
  {\pi(\Bgamma)q_r(\Bgamma^*|\Bchi_k^*)},
  \end{align*}
  proving  the algorithm has $\bar{\pi}(\cdot)$ as invariant distribution. Since this distribution has $\pi(\cdot)$ as marginal distribution it follows that $\Bgamma'\sim \pi(\cdot)$.
\end{proof}

Note that neither the large jump distribution $q_l(\cdot)$ nor the optimization  distribution $q_o(\cdot)$ (which can be both deterministic and stochastic)  are involved in the acceptance probability.  This gives great  flexibility  in the choice of these distributions.

Large jumps are not performed at each iteration, but rather through a composition of standard Metropolis-Hastings steps with local moves and large jumps. 
As a rule of thumb, based on suggestions of \citet{Tjelmeland99modejumping} and our own experience, we recommend that in not more than $5\%$ of the iterations large jumps are performed.
This is believed to provide a global Markov chain with both good mixing between the modes and accurate exploration of the regions around the modes. This in turn induces good performance of the algorithm in terms of the captured posterior mass for a given number of iterations. However, some tuning might well be required for the particular practical applications. 

The mode jumping MCMC steps can be modified to include a mixture of different proposal kernels $q_l, q_o$, and $q_r$ and parallelized using the multiple try MCMC idea. Technical details are given in~\ref{B}. 

\paragraph{An illustrative example} Assume 10 covariates $x_1,...,x_{10}$ and thus 1024 possible models. We generated $Y\sim N(1+10x_1+0.89x_8 + 1.43x_5,1)$ with correlated binary covariates (see supplementary code for details) and 1000 observations. We used a Gaussian linear regression with a Zellner's g-prior~\citep{zellner1986assessing} with $g = 1000$. This model has tractable marginal likelihoods described in detail in section~\ref{section4}. We consider an MJMCMC step with a large jump swapping randomly 4 components of $\Bgamma$ and a local greedy search, changing only one component at a time, as optimization routine. The last randomization changes each component of $\Bgamma$ independently with probability equal to 0.1. A typical MJMCMC step with locally optimized proposals is illustrated in Table~\ref{mjmcmcstep}.
\begin{table}[t]
\begin{adjustbox}{center}\begin{tabular}{ 
lrrrr}
\hline
&\multicolumn{2}{c}{\textbf{Forward}}&\multicolumn{2}{c}{\textbf{Backward}}\\\hline
&Model&$\log(p(\B y|\BB\gamma))$&Model&$\log(p(\B y|\BB\gamma))$\\\hline
Initial mode&$\Bgamma=$1010110111&1606.21&
$\Bgamma^*=$1101100001&1612.27\\
Large jump&$\Bchi_0^*=$10\textcolor{red}{01}110\textcolor{red}{00}1&1541.51&
$\Bchi_0=$11\textcolor{red}{10}100\textcolor{red}{11}1&1608.55\\
Optimize&$\Bchi_k^*=$1\textcolor{blue}{1}\textcolor{red}{01}1\textcolor{blue}{0}0\textcolor{red}{00}\textcolor{blue}{0}&1616.16&
$\Bchi_k=$1\textcolor{blue}{0}\textcolor{red}{10}100\textcolor{red}{11}\textcolor{blue}{0}&1612.00\\
Randomize&$\Bgamma^*=$1\textcolor{blue}{1}\textcolor{red}{01}1\textcolor{blue}{0}0\textcolor{red}{00}\textcolor{green}{1}&1612.27&
$\Bgamma=$1\textcolor{blue}{0}\textcolor{red}{10}1\textcolor{green}{1}0\textcolor{red}{11}\textcolor{green}{1}&1606.21\\
\hline
\multicolumn{5}{c}{Acceptance probability: $\min\left\{1,541.11\right\}$, accept $\Bgamma'=\Bgamma^*=$\textbf{1101100001}}\\\hline
\end{tabular}\end{adjustbox}
\\[1pt]
\caption{Illustration of a MJMCMC step with a large jump followed by a locally optimized proposal. The red components correspond to components swapped in the large jumps, the blue components to the ones changed in the optimizer, the green components of $\Bgamma$ to the randomization step.}\label{mjmcmcstep}
\end{table}

\paragraph{Large jumps} A change is defined by the components that  are to be swapped.  A simple  choice is to  give all  components an equal  probability $\rho$ to be swapped and independence between  components, in which case
\[
q_l(\Bchi_0^*|\Bgamma)=\prod_{j=1}^p\rho^{I_j}(1-\rho)^{1-I_j}
=\rho^{S}(1-\rho)^{p-S},
\]
where $I_j$ is a binary variable equal to  1 if  component $\gamma_j$ is to be swapped and $S=\sum_{j=1}^pI_j$ is the number of components to be swapped. 
An alternative is to 
first  draw the number of components, $S$, to swap according to a  distribution $q_S(\cdot)$ and thereafter  choose (uniformly) among the  possible changes of size $S$.
Table~\ref{proposals}
describes different ways of making large jumps where tuning parameters should be chosen such that the probability of a high value of $S$ is large.

\paragraph{Optimization}

In order to increase the quality of proposals and consequently both improve the acceptance ratio and increase the probability of escaping from local optima, the large jump is followed by a local optimization step.
Typically,  $q_o(\cdot)$ contains many iterations,  generating intermediate  states $\Bchi_0^*\rightarrow\Bchi_1^*\rightarrow\cdots\rightarrow\Bchi_k^*$ but none of these intermediate  states are needed for  the final evaluation. Different local learning and optimization routines can be applied for the generation of $\Bchi_k^*$, both deterministic and stochastic ones, see~\ref{app:prop} for further details. 
We will consider several feasible computationally options: local greedy optimization, local simulated annealing (SA) optimization, and local MCMC methods.

\paragraph{Randomization} A last randomization step defined through $q_r(\cdot) $ is needed in order to make the move back from $\Bgamma^*$ to $\Bgamma$ feasible.  We typically use randomizing kernels with a high mass on a small neighborhood around the mode but with a positive probability for any change. The two possible appropriate kernels from Table~\ref{proposals} are the random change of either random $S\sim \text{Unif}\{1,...,p\}$ or deterministic $S=p$ number of components with reasonably small but positive probabilities $0<\rho_i\ll 1$. This guarantees that the MJMCMC procedure is irreducible in $\Omega$.

\paragraph{Symmetric large jumps}
In order for the acceptance probability to be high, it is crucial that the auxiliary variables in the reverse sequence $\Bchi=(\Bchi_0,\Bchi_k)$ make $\Bgamma$ plausible ($q_r(\Bgamma|\Bchi_k)$ should be large in~\eqref{locmcmcgen00}). This may be difficult to achieve because the backwards large jump has no guarantee to be close to the current state. 
One way to achieve this is to choose $q_{l}(\Bchi^{*}_0|\Bgamma)$ to be symmetric, increasing the probability of returning close to the initial mode in the reverse large jump. The symmetry is achieved by swapping the same set of $\Bgamma_j$'s in the large jumps in the forward simulation as in the backwards simulation. 
We record the components $I$ that have been swapped.  
In our current implementation we require that only the components that do not correspond to $I$ can be changed in optimization transition kernels.
The following algorithm is a modification of Algorithm~\ref{MJMCMCalg0} taking a symmetric large jump into account.
\begin{algorithm}[H]
\caption{Mode jumping MCMC with symmetric backwards jump}\label{MJMCMCalg01}
\begin{algorithmic}[1]
\State  
Generate a large jump  $\Bchi_0^*$ by first generating a set $I\subset\{1,...,p\}\sim q_I(\cdot)$ defining the components to be swapped. 
\item Perform a local  optimization, defined through $\Bchi_k^*\sim q_o(\Bchi_k^*|\Bchi_0^*)$.
\State Perform a small randomization  to generate the proposal $\Bgamma^*\sim q_r(\Bgamma^*|\Bchi_k^*)$.
\item Define the backwards large jump $\Bchi_0$ through swapping the components $I$ in $\Bgamma^*$.
\item Generate  $\Bchi_k\sim q_o(\Bchi_k|\Bchi_0)$.
\State Put
\[
\Bgamma'=
\begin{cases}\Bgamma^*&\text{with probability $r_m(\Bgamma,\Bgamma^*;\Bchi_k,\Bchi_k^*)$;}\\
\Bgamma&\text{otherwise,}
\end{cases}
\]
where
\begin{equation}
r_{mh}^*(\Bgamma,\Bgamma^*;\Bchi_k,\Bchi_k^*) = \min\left\{1,\frac{\pi(\Bgamma^*)q_r(\Bgamma|\Bchi_{k})}{\pi(\Bgamma)q_r(\Bgamma^*|\Bchi_k^*)}\right\}\label{locmcmcgen01}.
\end{equation}
\end{algorithmic}
\end{algorithm}

\noindent
The following theorem  shows that also this  algorithm  also is $\pi$-invariant.
\begin{theorem}
Assume $\Bgamma\sim \pi(\cdot)$ and $\Bgamma'$ is generated according to   Algorithm~\ref{MJMCMCalg01}. 
Then  $\Bgamma'\sim \pi(\cdot)$.
\end{theorem}

\begin{proof}
The stochastic auxiliary components are now $I,\Bchi_k^*$ and $\Bchi_k$ where $\Bchi_0^*$ and
$\Bchi_0$ are deterministic functions of $(\Bgamma,I)$ and $(\Bgamma^*,I)$, respectively.
We have
\[
(\Bgamma,I,\Bchi_k^*)\sim \pi(\Bgamma)q_I(I)q_o(\Bchi_k^*|\Bchi_0^*)
\equiv \bar{\pi}(\Bgamma,I,\Bchi_k^*).
\]
We may now consider $(\Bgamma^*,I,\Bchi_k)$ as a proposal in the extended space, generated according to the distribution $q_r(\Bgamma^*|\Bchi_k^*)q_o(\Bchi_k|\Bchi_0)$.  An ordinary  Metropolis-Hastings iteration with respect to  $\bar{\pi}(\Bgamma,I,\Bchi_k^*)$ is then to accept $(\Bgamma^*,I,\Bchi_k)$ with probability
$r_{mh}^*=\min\{1,\alpha_{mh}^*\}$ where
\begin{align*}
\alpha_{mh}^*
=&\frac
  {\bar{\pi}(\Bgamma^*,I,\Bchi_k)q_r(\Bgamma|\Bchi_k)q_o(\Bchi_k^*|\Bchi_0^*)}
  {\bar{\pi}(\Bgamma,I,\Bchi_k^*)q_r(\Bgamma^*|\Bchi_k^*)q_o(\Bchi_k|\Bchi_0)}\\
  =&\frac
  {\pi(\Bgamma^*)q_I(I)q_o(\Bchi_k|\Bchi_0)
  q_r(\Bgamma|\Bchi_k)q_o(\Bchi_k^*|\Bchi_0^*)}
  {\pi(\Bgamma)q_I(I)q_o(\Bchi_k^*|\Bchi_0^*)
  q_r(\Bgamma^*|\Bchi_k^*)q_o(\Bchi_k|\Bchi_0)}
  =\frac
  {\pi(\Bgamma^*)q_r(\Bgamma|\Bchi_k)}
  {\pi(\Bgamma)q_r(\Bgamma^*|\Bchi_k^*)},
  \end{align*}
 proving  the algorithm has $\bar{\pi}(\cdot)$ as invariant distribution. Since this distribution has $\pi(\cdot)$ as marginal distribution it follows that $\Bgamma'\sim \pi(\cdot)$.
\end{proof}

\subsection{Delayed acceptance}

The most computationally demanding parts of the MJMCMC algorithms are the forward and backward optimizations. In many cases, the proposal generated through the forward optimization may lead to a very small value of $\pi(\Bgamma^*)$ resulting in a low acceptance probability regardless of the way the backwards auxiliary variables are generated. In such cases, one would like to reject directly without the need for performing the backward optimization. Such a scheme can be constructed by the use of the delayed acceptance procedure~\citep{christen2005markov,banterle2015accelerating}. We then have:
\begin{theorem}
Assume $\Bgamma\sim\pi(\cdot)$ and assume $\Bgamma^*$ is generated according to either Algorithm~\ref{MJMCMCalg0} or
Algorithm~\ref{MJMCMCalg01}. Accept $\Bgamma^*$ if
both
\begin{enumerate}
\item $\Bgamma^*$ is preliminary accepted with a probability $\min\{1,\tfrac{\pi(\Bgamma^*)}{\pi(\Bgamma)}\}$
\item and is finally accepted with a probability 
$\min\{1,\tfrac{q_r(\Bgamma|\Bchi_k)}{q_r(\Bgamma^*|\Bchi_k^*)}\}$.
\end{enumerate}
Then also $\Bgamma^*\sim\pi(\cdot)$.
\end{theorem}

\begin{proof}
We have that
\begin{align*}
\alpha_{mh}^*(\Bgamma,\Bgamma^*;\Bchi_k,\Bchi_k^*) 
=& \alpha_{mh}^1(\Bgamma,\Bgamma^*;\Bchi_k,\Bchi_k^*)\times\alpha_{mh}^2(\Bgamma,\Bgamma^*;\Bchi_k,\Bchi_k^*) 
\intertext{where}
\alpha_{mh}^1(\Bgamma,\Bgamma^*;\Bchi_k,\Bchi_k^*)=&\frac{\pi(\Bgamma^*)}{\pi(\Bgamma)},\quad
\alpha_{mh}^2(\Bgamma,\Bgamma^*;\Bchi_k,\Bchi_k^*)=\frac{q_r(\Bgamma|\Bchi_{k})}{q_r(\Bgamma^*|\Bchi_k^*)}.
\end{align*}
Since $\alpha_{mh}^j(\Bgamma,\Bgamma^*;\Bchi_k,\Bchi_k^*)
=[\alpha_{mh}^j(\Bgamma^*,\Bgamma;\Bchi_k^*,\Bchi_k)]^{-1}$ for $j=1,2$, it follows by the general results in~\citet{banterle2015accelerating} that we obtain an invariant kernel with respect to $\bar{\pi}$.
\end{proof}

In general the total acceptance rate will be smaller than without delayed acceptance~\citep[][remark 1]{banterle2015accelerating}, but the gain by avoiding a backwards optimization step if not accepted in the preliminary step can compensate on this.

\subsection{Calculation of marginal densities}\label{sec:marg.dens}

In practice exact calculation of the marginal density can only be performed in simple models such as linear Gaussian ones, so alternatives need to be considered. 
One approach is to use estimators that are accurate enough to neglect the approximation errors involved. Such approximative approaches have been used in various settings of Bayesian variable selection and Bayesian model averaging. Laplace's method \citep{tierney1986accurate} has been widely used, but is based on rather strong assumptions. The harmonic mean estimator~\citep{newton1994approximate} is an easy to implement MCMC based method but can give high variability in the estimates.
Chib's method~\citep{chib1995marginal}, and its extension~\citep{chib2001marginal}, 
have gained increasing popularity and can be very accurate provided enough MCMC iterations are performed. Approximate Bayesian Computation~\citep{marin2012approximate} has also been considered in this context, being much faster than MCMC alternatives, but also giving cruder approximations.
Variational methods~\citep{jordan1999introduction} provide lower bounds for the marginal likelihoods and have been used for model selection in e.g. mixture models~\citep{mcgrory2007variational}. 
Integrated nested Laplace approximation~\citep[INLA,][]{rue2009eINLA} provides accurate estimates of marginal likelihoods within the class of latent Gaussian models. 
In the context of generalized linear models, BIC type approximations can be used.

An alternative is to insert unbiased estimates  of $\pi(\Bgamma)$ into the Metropolis-Hastings acceptance probabilities. \citet{andrieu2009pseudo} name this the
\emph{pseudo-marginal} approach and show that this leads to exact algorithms (in the sense of converging to the right distribution).  Importance sampling ~\citep{beaumont2003} and particle filter~\citep{andrieu2010particle} are two approaches that can be used within this setting. 
In general, the convergence rate will depend on the amount of Monte Carlo effort that is applied. \citet{doucet2015efficient} provide some guidelines.

Our implementation of the MJMCMC algorithm allows for all of the available possibilities for calculation of marginal likelihoods and assumes that the approximation error can be neglected. For the experiments in section~\ref{section4} we have applied exact evaluations in the case of linear Gaussian models, approximations  based on the assumed informative priors in case of generalized linear models \citep{Clyde:Ghosh:Littman:2010}, and INLA \citep{rue2009eINLA} in the case of latent Gaussian models. \citet{bivand2015spatial} also apply INLA within an MCMC setting, but then concentrating on hyperparameters that (currently) can not be estimated within the INLA framework. 
\citet{Friel2012} performed comparison of some of the mentioned approaches for calculation of marginal likelihoods, including  Laplace's  approximations,  harmonic  mean  approximations,  Chib's  method  and  others. \citet{HubinStorvikINLA} reported some comparisons of INLA and other methods for approximating marginal likelihood. There it is demonstrated that INLA provides extremely accurate approximations on marginal likelihoods in a fraction of time compared to Monte Carlo based methods.  \citet{HubinStorvikINLA} also demonstrated that by means of adjusting tuning parameters within the algorithm \citep[the grid size and threshold values within the numerical integration procedure, ][]{rue2009eINLA} one can often make the difference between INLA and unbiased methods of estimating of the marginal likelihood arbitrary small.

\subsection{Parallelization and tuning parameters of the search}

With large number of potential explanatory variables it is important to be able to utilize multiple cores and GPUs of either local machines or clusters in parallel.
General principles of utilizing multiple cores in local optimization are provided in~\citet{multcpuopt}. 
At every step of the local optimization within the large jump steps we allow to simultaneously draw several proposals with respect to a certain transition kernel during the optimization procedure and then sequentially calculate the transition probabilities as the proposed models are evaluated by the corresponding CPUs, GPUs or clusters in the order they are returned. In those iterations where no large jumps are performed, we are utilizing multiple cores by means of addressing multiple try MCMC to explore the solutions around the current mode. The parallelization strategies are described in detail in~\ref{B}.

In practice, tuning parameters of the local optimization routines such as the choice of the neighborhood, generation of proposals within it, the cooling schedule for \textit{simulated annealing} \citep{Michiels2005optim} or number of steps in greedy optimization also become crucially important and it yet remains unclear whether we can optimally tune them before or during the search. Mixing of proposals from Table~\ref{proposals} and of optimizers is also possible.
Tuning the probabilities of addressing these different options can be  beneficial. Such tuning is a sophisticated mathematical problem, which we are not trying to resolve optimally within this paper, however we suggest a simple practical idea for obtaining reasonable solutions. Within the BAS algorithm, an important feature was to utilize the marginal inclusion probabilities of different covariates. We have introduced this in our algorithms as well by allowing insertion of estimates of the $\rho_i$'s in proposals given in Table~\ref{proposals} based on some burn-in period. They then correspond to the marginal inclusion probabilities after burn-in shifted with some small $\epsilon$ from 0 and 1 if necessary in order to guarantee irreducibility. Additional literature review on search parameter tuning can be found in \citet{GangLuo}.

\section{Experiments}\label{section4}
In this section we are going to apply the MJMCMC algorithm to different data sets and analyze the results in relation to other algorithms. Linear regression is addressed through  
the U.S. Crime Data \citep{doi:10.1080/01621459.1997.10473615} and a protein activity data~\citep{CLYDE01061998}. 
Logistic regression is considered in 
a simulated example based on a data set  and
through an Arabidopsis epigenetic data set. The Arabidopsis example also includes random effects. 

We compare the performance of our approach to competing MCMC methods such as the MCMC model composition algorithm~\citep[$\text{MC}^3$,][]{madigan1995bayesian, doi:10.1080/01621459.1997.10473615} and the random-swap (RS) algorithm~\citep{Clyde:Ghosh:Littman:2010} as
well as the BAS algorithm~\citep{Clyde:Ghosh:Littman:2010}.
Both $\text{MC}^3$ and RS are simple MCMC procedures based on the standard Metropolis-Hastings algorithm with proposals chosen correspondingly as an inversion or a random change of one coordinate in $\Bgamma$ at a time  \citep{Clyde:Ghosh:Littman:2010}.
BAS carries out sampling without repetition from the space of models with respect to the adaptively updated marginal inclusion probabilities.  For one of the examples, also a comparison with the ESS++ software~\citep[evolutionary stochastic search,][]{Bottolo15022011} is made. 
For the cases when full enumeration of the model space is possible we additionally compare all of the aforementioned approaches to the benchmark TOP method that consists of the best quantile of models in terms of the posterior probability for the corresponding number of addressed models $\Vert\mathbb{V}\Vert$ and can not by any chance be outperformed in terms of the posterior mass captured. 

The different algorithms that are compared are implemented in different programming languages, making it difficult to compare CPU time fairly. We have therefore focused on both the total number of visited models and the number of unique models visited, since this is the main computational burden (marginal likelihood values of visited models can be stored). The number of models visited for MJMCMC includes all of the models visited during global and local moves as well as local combinatorial optimization, hence the comparison on the same number of totally visited and uniquely visited models is fair.

Following \citet{Clyde:Ghosh:Littman:2010}, approximations for model probabilities \eqref{approxpost} and marginal inclusion probabilities \eqref{margininuspost} based on a subspace of models are further referred to as RM (renormalized) approximations, whilst the corresponding MCMC based approximations \eqref{map2} 
are referred to as MC approximations. The validation criteria addressed include  root mean squared errors and bias of parameters of interest based on multiple replications of each algorithm, similar 
to~\citet{Clyde:Ghosh:Littman:2010}. In addition to marginal
inclusion probabilities, we also include a global measure
\begin{align}
 \text{C}(\Bgamma)
 =&\frac{\sum_{\Bgamma' \in \mathbb{V}} p(\By| \Bgamma')p(\Bgamma')}
   {\sum_{\Bgamma' \in\Omega} p(\By| \Bgamma')p(\Bgamma')},\label{def.C}
\end{align}
describing the fraction of probability mass contained in the subspace $\mathbb{V}$. This measure allows us  to address how well the search works in terms of capturing posterior mass within a given model space. By formula \eqref{approxpost} maximization of $\text{C}(\Bgamma)$ automatically induces minimization of the bias in terms of posterior marginal model probabilities, which vanishes gradually when $\text{C}(\Bgamma)\rightarrow 1$.

Mixtures of different proposals from Table~\ref{proposals} and local optimizers mentioned in section~\ref{mjmcmcsimp} were used in the studied examples in the MJMCMC algorithm. A validation of the gain in using such mixtures is given in example~\ref{expir2}, where we address both MJMCMC with mixtures and a simpler version where only one choice of proposal distributions is used (the details are given in the example). The details on the choices and frequencies of different proposals for the other examples are given  in Tables~\ref{optimparam}-\ref{proposals5}
in~\ref{appccc}. The choices are based on some tuning on a simulated data example, reported in section~\ref{sec:example.s.1}. Further small adaptations were made in some of the examples. Generally speaking, we can not claim that the choices of the tuning parameters are optimal. It is rather some subjectively rational choice.  

\subsection{Example 1}\label{expir2}

Here we address the U.S. Crime data set, first introduced by \citet{Vandaele1978} and stated to be a test bed for evaluation of methods for model selection~\citep{doi:10.1080/01621459.1997.10473615}. The data set consists of $n=47$ observations on 15 covariates and the responses, which are the corresponding crime rates. We will compare performance of the algorithms based on a linear Bayesian regression model using a Zellner's g-prior \citep{zellner1986assessing} with $g = 47$. This implies that the marginal likelihood is of the following form: 
\begin{equation}
p(\By|\Bgamma) \propto (1+g)^{(n-p-1)/2}(1+g[1-R_{\Bgamma}^2])^{-(n-1)/2},\label{gelmlik}
\end{equation}
where $R_{\Bgamma}^2$ is the usual coefficient of determination of a linear regression model. With this scaling, the marginal likelihood of the null model (the model containing no covariates) is 1.0.

This is a sophisticated example with a total of $2^{15}=32768$ potential models and with several local modes. As a result, all simple MCMC methods easily get stuck and have extremely poor performances in terms of the captured mass and precision of both the marginal posterior inclusion probabilities and the posterior model probabilities.
Table~\ref{rmse.sim.2} shows the RMSE (scaled by $10^2$) for the model parameters over 100 repeated runs for each algorithm. The True column contains the true marginal inclusion probabilities (obtained from full enumeration) while the TOP column shows the RMSE results based on the 3276 models with \emph{highest} posterior probabilities (about 10\% of the total number of models). The MJMCMC columns show the results based on using mixtures of proposals  and optimizers (see Tables~\ref{optimparam} and~\ref{proposals1}  for details) while the MJMCMC$^*$ results are based on one specific choice of proposals with swaps of 4 components at a time for the large jumps (Type 4 in Table~\ref{proposals}) and a local greedy optimizer changing two components at a time with a last randomization of type 2 (Table~\ref{proposals}). For the standard MCMC steps, a type 4 with two changing components was used.  

For this example, both the MC$^3$ and the RS methods  got stuck in some local modes and for the 3276 models only 829/1071 unique models where visited. These algorithms did not reach 3276 unique models within a reasonable time for this example (most likely the algorithm could not escape from local extrema), hence such a scenario is not reported.
For this example MJMCMC gives a much better performance than the other MCMC methods in terms of both MC and RM based estimations with respect to the posterior mass captured, $\text{C}(\BB\gamma)$. With a total of 3276 visited, BAS slightly outperforms MJMCMC. However, when running MJMCMC so that the number of \emph{unique} models visited ($\Vert\mathbb{V}\Vert$) are comparable with BAS, MJMCMC gives better results (columns marked with MJMCMC$^2$ in Table~\ref{rmse.sim.2}).  The comparison is performed in terms of posterior mass captured, biases and root mean squared errors for both posterior model probabilities and marginal inclusion probabilities (Table~\ref{rmse.sim.2}).

BAS has the property of always visiting new unique models, whilst all MCMC based procedures tend to do revisiting with respect to the corresponding posterior probabilities. When generating a proposal is much cheaper than estimating marginal likelihoods of the model (which is usually the case, also in this example) and we are storing the results for the already visited models, having generated a bit more models by MJMCMC does not seem to be a serious issue. Those unique models that are visited have a higher posterior mass than those suggested by BAS (for the same number of models visited). Furthermore MJMCMC (like BAS) can escape from local modes. 

Also the results based on no mixture of proposals (MJMCMC* in the table)
are much better than standard MCMC methods, however the results obtained by the MJMCMC algorithm with a mixture of proposals were even better. 
We have tested this on some other examples too and the use of mixtures was always beneficial and thus recommended. For this reason only the cases with mixtures of proposals are addressed in other experiments.

\begin{table}[!t]
\resizebox{\textwidth}{!}{%
\begin{tabular}{crrrrrrrHrrrrHrHr}
\hline
Par&True&TOP&\multicolumn{2}{c}{MJMCMC}&\multicolumn{2}{c}{MJMCMC$^2$}&\multicolumn{1}{c}{BAS}&&\multicolumn{2}{c}{$\text{MC}^3$}&\multicolumn{2}{c}{RS}&\multicolumn{4}{c}{MJMCMC*}\\
\hline 
$\Delta$&$\pi_j$&-&RM&MC&RM&MC&RM&RM&MC&RM&MC&RM&RM&RM&MC&MC\\\hline
$\gamma_{8}$&0.16&3.51&6.57&10.68&5.11&10.29&5.21&6.36&6.49&3.49&5.87&3.31&-6.19&6.23&-7.74&9.06\\
$\gamma_{13}$&0.16&3.34&7.46&10.54&5.60&10.19&6.26&7.20&8.62&3.39&8.83&3.05&-6.31&6.38&-5.45&10.54\\
$\gamma_{14}$&0.19&3.24&8.30&12.43&6.30&12.33&6.20&7.39&6.58&2.55&6.22&2.46&-7.10&7.15&-7.77&10.91\\
$\gamma_{12}$&0.22&3.27&6.87&13.61&5.57&13.64&3.10&5.21&5.81&6.23&4.93&5.27&-5.14&5.29&-3.30&10.93\\
$\gamma_{5}$&0.23&2.56&6.30&13.45&4.59&13.65&1.84&4.02&6.07&13.05&5.13&12.77&-5.22&5.39&-9.01&10.90\\
$\gamma_{9}$&0.23&3.27&9.49&16.21&7.40&16.21&9.27&8.37&5.99&2.99&5.70&2.60&-7.58&7.68&-8.54&11.06\\
$\gamma_{7}$&0.29&2.31&4.37&13.63&3.45&12.73&2.28&2.96&4.74&9.61&3.46&9.70&-3.64&3.91&-5.81&10.10\\
$\gamma_{4}$&0.30&1.57&6.18&19.22&3.79&17.31&0.99&1.20&13.24&21.84&13.53&21.48&-4.12&4.63&-9.60&13.22\\
$\gamma_{6}$&0.33&1.92&8.61&19.71&6.14&19.49&3.11&5.30&10.19&7.47&10.99&7.12&-5.59&5.87&-12.00&15.43\\
$\gamma_{1}$&0.34&2.51&11.32&22.68&7.29&20.50&8.43&7.20&22.89&25.19&23.63&24.71&-7.35&7.58&-9.15&12.97\\
$\gamma_{3}$&0.39&0.43&3.95&11.13&2.38&6.99&5.02&3.78&21.48&30.24&21.39&29.94&1.86&2.99&-7.62&12.66\\
$\gamma_{2}$&0.57&1.58&5.92&13.21&3.82&9.03&13.78&8.66&30.81&37.57&29.27&37.15&4.57&5.11&-10.22&14.04\\
$\gamma_{11}$&0.59&0.58&3.57&13.49&2.37&15.94&4.04&2.18&11.88&21.79&11.16&21.31&1.61&2.77&-5.58&12.77\\
$\gamma_{10}$&0.77&3.25&7.62&7.28&5.97&4.78&15.45&10.46&21.83&19.18&20.53&19.65&6.25&6.41&-10.59&14.27\\
$\gamma_{15}$&0.82&3.48&9.23&4.45&6.89&5.85&14.50&11.75&69.68&76.81&69.19&76.30&6.59&6.75&-10.72&14.76\\\hline
$\text{C}(\Bgamma)$&1.00&0.86&0.58&0.58&0.71&0.71&0.66&0.67&0.10&0.10&0.10&0.10&0.60&0.60&0.60&0.60\\\hline
Eff&$2^{15}$&3276&1909&1909&3237&3237&3276&3276&829&829&1071&1071&3264&3264&3264&3264\\\hline
Tot&$2^{15}$&3276&3276&3276&5936&5936&3276&3276&3276&3276&3276&3276&4295&4295&4295&4295\\\hline
\end{tabular}}
\\[1pt]
\caption{\label{rmse.sim.2}Average root mean squared error (RMSE) over the 100 repeated runs of every algorithm on the Crime data [example 1); the values reported in
the table are RMSE $\times 10^2$ for $p(\gamma_j=1|\By)$. $\text{C}(\Bgamma)$ is defined in \eqref{def.C}.
Tot is the total number of visited models, while Eff is the number of unique models visited during the iterations of the algorithms (for the TOP column all $2^{15}$ models were visited but the RMSE are based on the best 3276 models). RM corresponds to using the renormalization procedure~\eqref{approxpost} while MC corresponds to using the MC procedure~\eqref{map2}. 
MJMCMC$^2$ differ from MJMCMC in the number of unique models  visited (Eff) while MJMCMC$^*$ corresponds to a run with no mixtures of proposals.
The corresponding biases are reported in~\ref{sec:further.results} in Table~\ref{bias.sim.2}.}
\end{table}

\subsection{Example 2}\label{sec:example.s.2}
\begin{figure}[t]
    \centering
    \includegraphics[width=0.5\textwidth]{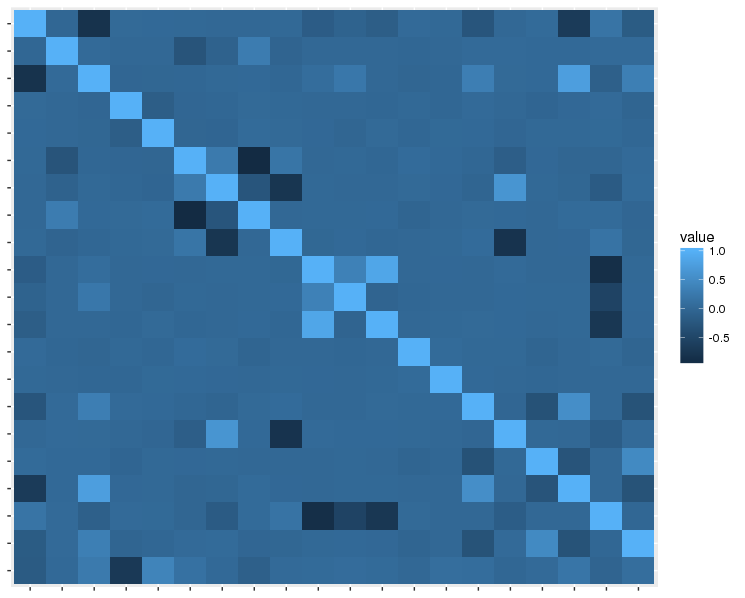}
    \caption{Correlation structure of the covariates in example 3.}
    \label{ex3corr}
\end{figure}

In this example we are considering a new simulated data set for logistic regression. We generated $p=20$ covariates as a mixture of binary and continuous variables. The correlation structure is shown in Figure~\ref{ex3corr} while the full details of how the data was generated is given in~\ref{C4}. 

A total of $2^{20}=1048576$ potential models need to be considered in this case. Additionally, in this example $n = 2000$, which makes estimation of a single model significantly slower than in the previous example. 
For $\BB\gamma$ we use the binomial prior~\eqref{glmgammaprior} with $q=0.057$.
We are in this case using the BIC-approximation for the marginal likelihood,
\begin{equation}
\log \widehat{p}(\mathbf{y}|\Bgamma) = \log\hat p(\B y|\hat{\BB\beta}_{\BB\gamma}) -\frac{n}{2}\log(|\hat{\BB\beta}_{\BB\gamma}|),\label{ex3mlik}
\end{equation}
where $\hat{\BB\beta}_{\BB\gamma}$ is the maximum likelihood (or MAP) estimate
for the $\beta_j$'s involved and $|\hat{\BB\beta}_{\BB\gamma}|$ is the number of parameters.  This choice was made in order to compare the results with implementations of BAS, RS and MC$^3$ available in the supplementary to~\citet{Clyde:Ghosh:Littman:2010}, where this approximation is considered. In that way, the model search procedures are compared based on the same selection criterion.

\begin{table}[!t]
\resizebox{\textwidth}{!}{%
\begin{tabular}{ 
crrrrrrrcrrr}
\hline

Par&True&TOP&\multicolumn{2}{c}{MJMCMC}&\multicolumn{2}{c}{MJMCMC$^2$}&BAS&BAS-RS&\multicolumn{2}{c}{RS}\\
\hline 
$\Delta$&$\pi_j$&-&RM&MC&RM&MC&RM&RM&RM&MC\\ \hline
 $\gamma_{6}$&0.29&0.00&7.38&15.54&4.54&16.62&6.47&3.67&6.01&2.11\\ 
 $\gamma_{8}$&0.31&0.00&6.23&15.50&3.96&16.94&5.58&3.02&5.37&2.55\\ 
 $\gamma_{12}$&0.35&0.00&4.86&14.62&2.78&13.66&4.22&2.12&3.91&2.37\\ 
 $\gamma_{15}$&0.35&0.00&4.55&15.24&2.56&15.45&4.66&1.64&3.40&2.56\\ 
 $\gamma_{2}$&0.36&0.00&4.90&16.52&2.92&17.39&5.42&2.45&3.65&2.61\\ 
 $\gamma_{20}$&0.37&0.00&4.82&14.35&2.66&14.08&3.32&1.80&4.15&2.18\\ 
 $\gamma_{3}$&0.40&0.00&9.25&20.93&5.65&22.18&9.75&4.82&6.76&2.83\\ 
 $\gamma_{14}$&0.44&0.00&3.14&17.54&1.58&16.24&3.73&1.30&1.33&2.93\\ 
 $\gamma_{10}$&0.44&0.00&4.60&18.73&2.29&17.90&4.87&1.30&1.51&2.42\\ 
 $\gamma_{5}$&0.46&0.00&3.10&17.17&1.53&16.97&4.06&1.51&1.09&2.85\\ 
 $\gamma_{9}$&0.61&0.00&3.68&16.29&1.63&13.66&3.89&1.39&2.19&2.35\\ 
 $\gamma_{4}$&0.88&0.00&5.66&6.70&3.74&6.26&6.60&5.57&7.61&2.15\\ 
 $\gamma_{11}$&0.91&0.00&5.46&6.81&3.95&6.90&4.66&3.14&4.32&1.57\\ 
 $\gamma_{1}$&0.97&0.00&1.90&1.74&1.35&1.34&2.43&1.96&2.30&1.1\\ 
 $\gamma_{13}$&1.00&0.00&0.00&0.43&0.00&0.32&0.00&0.00&0.00&0.37\\ 
 $\gamma_{7}$&1.00&0.00&0.00&0.57&0.00&0.41&0.00&0.00&0.00&0.33\\ 
 $\gamma_{16}$&1.00&0.00&0.00&0.41&0.00&0.33&0.00&0.00&0.00&0.23\\ 
 $\gamma_{17}$&1.00&0.00&0.00&0.43&0.00&0.39&0.00&0.00&0.00&0.23\\ 
 $\gamma_{18}$&1.00&0.00&0.00&0.47&0.00&0.35&0.00&0.00&0.00&0.24\\ 
 $\gamma_{19}$&1.00&0.00&0.00&0.52&0.00&0.36&0.00&0.00&0.00&0.41\\ 
\hline $\text{C}(\BB\gamma)$&1.00&1.00&0.72&0.72&0.85&0.85&0.74&0.85&0.68&0.68\\ 
\hline Eff&$2^{20}$&10000&5148&5148&9988&9988&10000&10000&1889&1889\\ 
\hline Tot&$2^{20}$&10000&9998&9998&19849&19849&10000&10000&10000&10000\\ \hline

\end{tabular}}
\\[1pt]
\caption{Average root mean squared error (RMSE) from the 100 repeated runs of every algorithm on the simulated logistic regression data (example 2); the values reported in
the table are RMSE $\times 10^2$ for $p(\gamma_j=1|\By)$. See the caption of Table~\ref{rmse.sim.2} for further details. The corresponding biases are reported in the appendix \ref{sec:further.results} in Table~\ref{bias.sim.3}.}\label{rmse.sim.3}
\end{table}

Some of the covariates involved have large correlations.
This induces both multimodality within the space of models and sparsity of the locations of the modes and creates an interesting example for comparison of different search strategies. As one can see in Table~\ref{rmse.sim.3}, MJMCMC outperformed pure BAS by far both in terms of posterior mass captured and in terms of root mean square errors of marginal inclusion probabilities
when based on the same number of unique models. MJMCMC 
outperformed RS as well. The latter got stuck in some local modes and could only reach 1889 unique models for the 10000 models visited. We could not reach 10000 unique models for the RS algorithm within a reasonable time for this example either (again most likely the algorithm could not escape from local extrema), hence such a scenario is not reported.  Even for almost two times less originally visited models in $\mathbb{V}$, comparing to BAS, MJMCMC gives almost the same results in terms of the posterior mass captured and errors. MJMCMC, for the given number of unique models visited, did not outperform a combination of MCMC and BAS (BAS-RS), which is recommended by~\citet{Clyde:Ghosh:Littman:2010} for larger model spaces; both of them gave approximately identical results.

\subsection{Example 3}
This experiment is based on a much larger model space in comparison to all of the other examples. We address the protein activity data \citep{CLYDE01061998} and consider all main effects together with the two-way interactions and quadratic terms of the continuous covariates resulting in 88 covariates in total. This corresponds to a model space of cardinality $2^{88}$, a number far to high to perform full search through all models. This model space is additionally multimodal, which is the result of having high correlations between numerous of the addressed covariates (17 pairs of covariates have correlations above 0.95).
\begin{figure}[!t]
\centering
\includegraphics[width=0.45\linewidth]{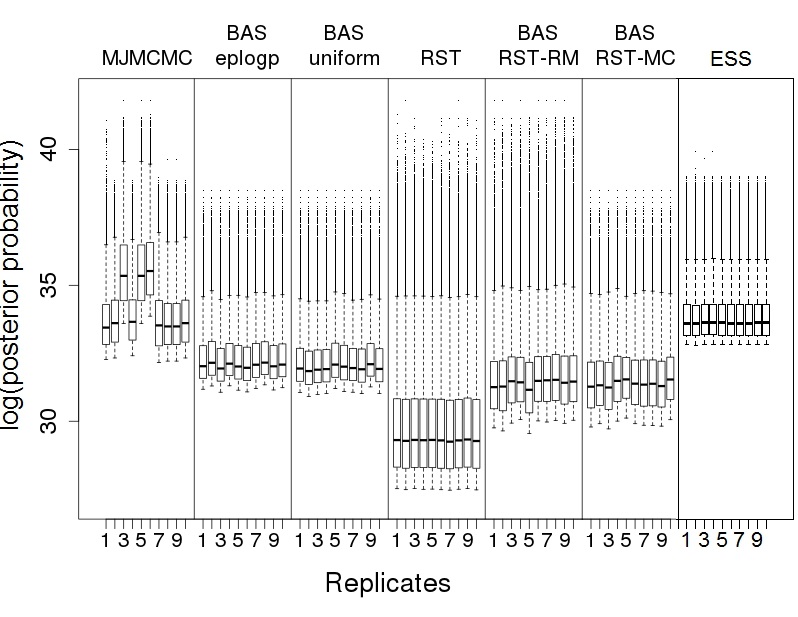}
\includegraphics[width=0.51\linewidth]{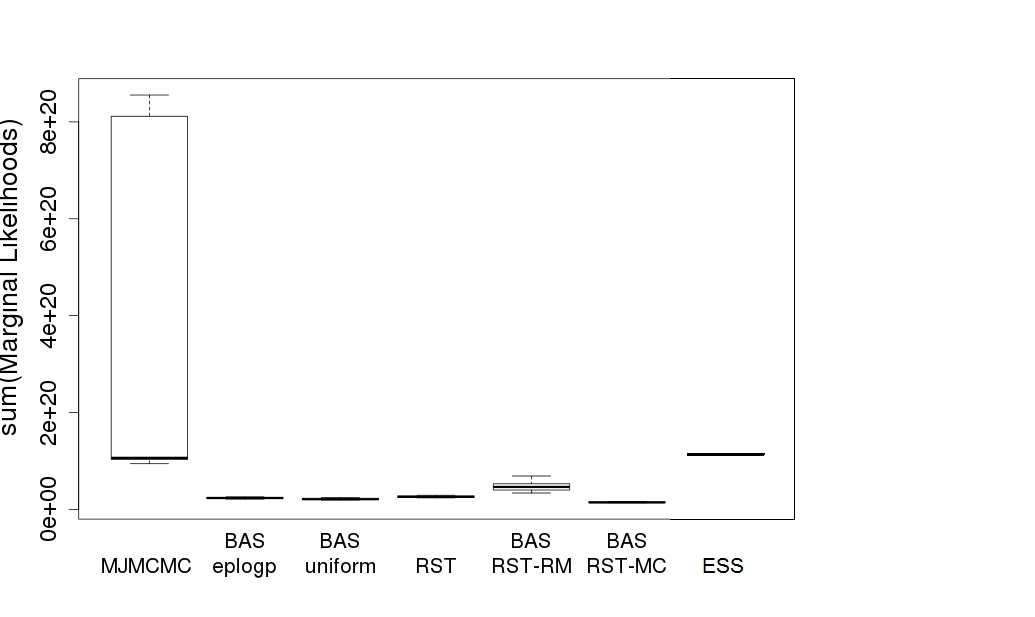}
\caption{Comparisons   in the protein data of the log posterior probabilities of the top 100000 models (left) and box-plots of the posterior mass captured (right) obtained by MJMCMC, BAS-eplogp, BAS-uniform, thinned version of Random Swap (RST), BAS with Monte Carlo estimates of inclusion probabilities from the RST samples (BAS-RST-MC), BAS re-normalized estimates of inclusion probabilities (BAS-RST-RM) from the RST samples, and ESS++ (ESS).}\label{postincusion}
\end{figure}
We analyzed the data set using Bayesian linear regression with the binomial prior~\eqref{glmgammaprior} with $q=0.5$ for $\BB\gamma$ and a Zellner's g-prior with $g = 96$ for $\BB\beta$ (the data has $n=96$ observations). 
We then compared the performance of MJMCMC, BAS and RS. For this example we have also addressed the ESS++ algorithm \citep{Bottolo15022011}.

The reported RS results are based on the RS  algorithm run for $88\times 2^{20}$ iterations and a thinning rate of $\frac{1}{88}$ (named RST in \citet{Clyde:Ghosh:Littman:2010}). BAS was run with several choices of initial sampling probabilities such as uniformly distributed within the model space one, eplogp  adjusted~\citep{Clyde:Ghosh:Littman:2010}, and those based on RM and MC approximations obtained by the RST algorithm. For the first two initial sampling probabilities BAS was run for $2^{20}$ iterations. For the two latter (the BAS-RST-RM and BAS-RST-MC) algorithms) first RS was run for $88\times 2^{19}$ iterations providing $2^{19}$ models for estimating initial sampling probabilities and then BAS was run for the other $2^{19}$ iterations based on RM or MC estimates of the marginal inclusion probabilities. MJMCMC was run until $2^{20}$ unique models were obtained. ESS++ was run with default search settings until $2^{20}$ unique models were visited. All of the algorithms were replicated 10 times. 

\begin{figure}[t]
\centering
\includegraphics[width=0.48\linewidth]{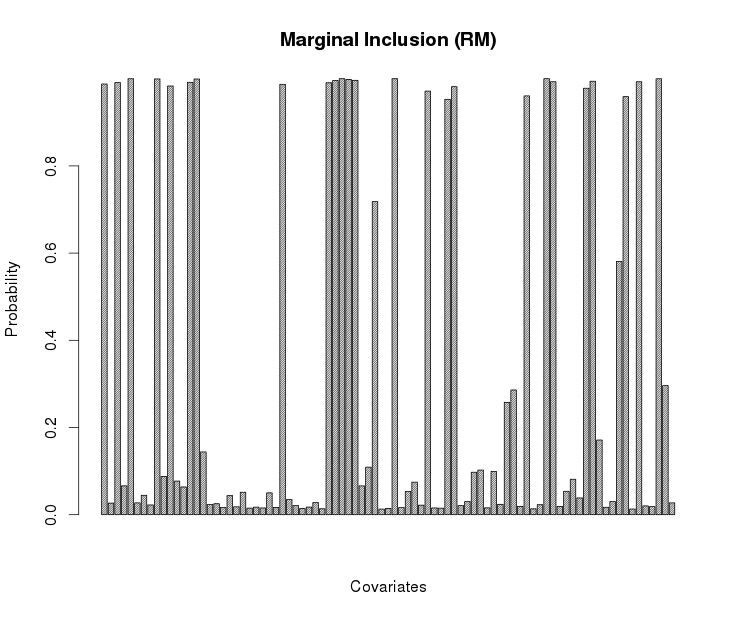}
\includegraphics[width=0.48\linewidth]{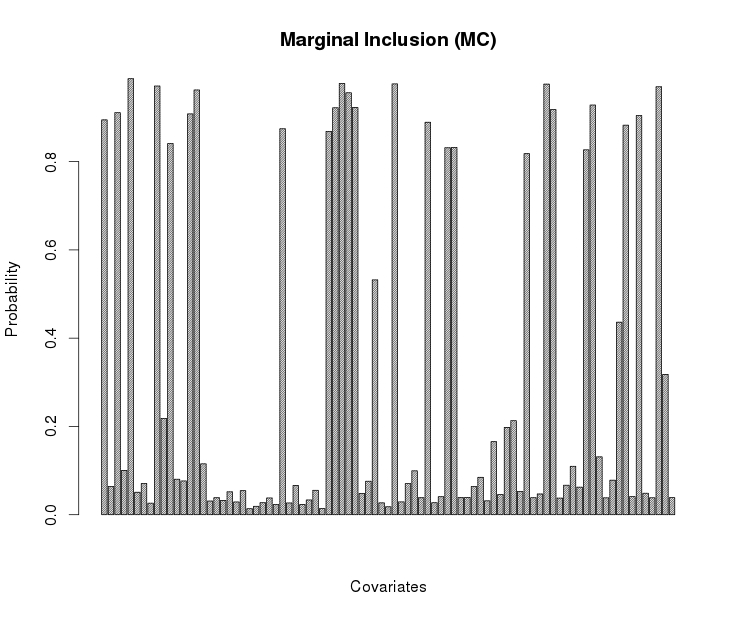}
\caption{Comparisons of RM (left) and MC (right) estimates of marginal posterior inclusion probabilities obtained by the best run of MJMCMC with $8.56\text{e}+20$ posterior mass captured.}\label{figmlikmass}
\end{figure}
In Figure~\ref{postincusion} box-plots of the best 100000 models captured by the corresponding replications of the algorithms as well as posterior masses captured by them are displayed. 
BAS with both uniform and eplogp initial sampling probabilities performed rather poorly in comparison to other methods, whilst BAS combined with RM approximations from RST did slightly better. ESS++ as well as MJMCMC show the most promising results. BAS with RM initial sampling probabilities usually managed to find models with the highest posterior probabilities, however MJMCMC in general captured by far higher posterior mass within the same amount of unique models addressed. Marginal inclusion probabilities obtained by the best run of MJMCMC with respect to mass (denominator of~\eqref{approxpost} with value $8.56\times 10^{20}$ in Figure~\ref{postincusion}) are reported in Figure~\ref{figmlikmass}, whilst those obtained by other methods can be found in \citet{Clyde:Ghosh:Littman:2010}. Since MJMCMC obtained the highest posterior mass, we expect that the corresponding RM estimates of the marginal inclusion probabilities are the least biased, moreover they perfectly agree with the MC approximations. Although MJMCMC in all of the obtained replications outperformed most of the competitors in terms of the posterior mass captured, it itself exhibited significant variation between the runs (right panel of Figure~\ref{postincusion}). The latter issue can be explained by that we are only allowing visiting $3.39\times 10^{-19}\%$ of the total model space in the addressed replications, which might be not enough to always converge to the same posterior mass captured. Note however that the variability in the results obtained from different runs of MJMCMC clearly indicates that more iterations are needed, while the other methods may indicate (wrongly) that sufficient iterations have been performed.

\subsection{Example 4}
In this example we illustrate how MJMCMC works for GLMM models. As illustration, we address genomic and epigenomic data on Arabadopsis.
Arabadopsis is a plant model organism with a lot of genomic/epigenomic data easily available \citep{becker2011spontaneous}. At each position on the genome, a number of reads are allocated. At locations with a nucleotide of type cytosine (C), reads are either methylated or not. Our focus will be  on  modeling the amount of methylated reads through different covariates including (local) genomic structures, gene classes and expression levels. The studied data was obtained from the NCBI GEO archive \citep{barrett2013ncbi}.

We model the number of methylated reads $Y_i\in \{1,...,R_i\}$ per loci $i = 1,...,n$, where $R_i \in \mathbb{N}$ is the number of reads, through~\eqref{themodeleq}-\eqref{themodeleqend} by a Poisson distribution for the response and $n = 1502$. Since in general the ratio of methylated bases is low, we have preferred the Poisson distribution of the responses to the binomial. 
The mean $\eta_i$ is  modeled via the log link to the chosen covariates, including an offset defined by  $R_i$ per location, and a spatially correlated  random effect $\delta_i$ which is modeled via an $AR(1)$ process with parameter $\rho \in \mathbb{R}$ , namely $ \delta_i =  \rho\delta_{i-1} + \epsilon_i\in \mathbb{R}$ with $\epsilon_{i} \sim N(0,\sigma_{\epsilon}^2)$, $i=1,...,n$. Thus, we take into account spatial dependence structures of methylation rates along the genome as well as the variance of the observations not explained by the covariates.  
We use the binomial prior~\eqref{glmgammaprior} with $q=0.5$ for $\Bgamma$ and the Gaussian prior for the regression coefficients:
\[
\boldsymbol{\beta}|\Bgamma \sim  N_{p_{\Bgamma}}(\boldsymbol{\mu_{\beta_\gamma}},\boldsymbol{\Sigma_{{\beta_\gamma}}}).
\]
For the parameters within the random effects, we first reparametrize to $\psi_1 = \log{\frac{1}{\sigma_{\epsilon,t}^2}(1-\rho^2)}$, $\psi_2 =   \log{\frac{1+\rho}{1-\rho}}$ and assume
\begin{align}
\psi_1 \sim& \text{logGamma}(1,5\times 10^{-5})\\
\intertext{and}
\psi_2 \sim& N(0,0.15^{-1}).
\end{align}
Marginal likelihoods were for this example calculated through the INLA package  (\url{www.r-inla.org}).
\begin{figure}[t]
    \centering
    \includegraphics[width=0.5\textwidth]{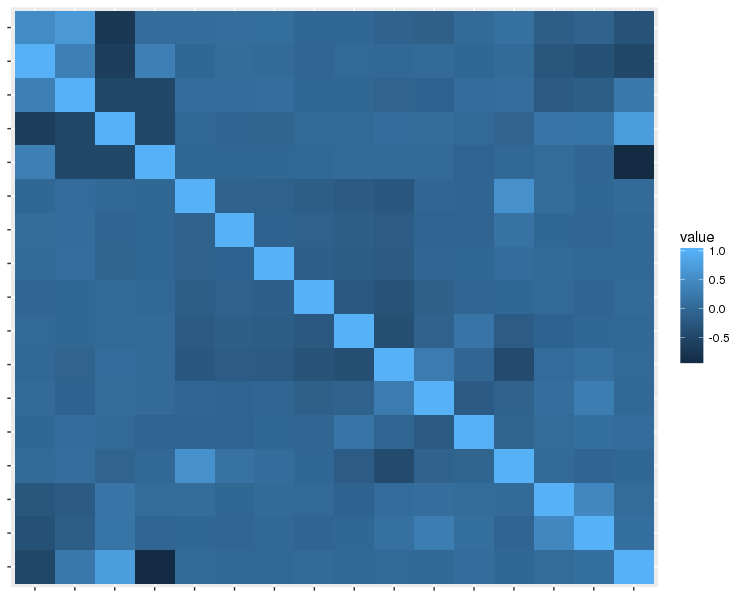}
    \caption{Correlation structure of the covariates in example~4.}
    \label{ex5corr}
\end{figure}
\begin{table}[!t]
\begin{adjustbox}{center}
\begin{tabular}
{crHrHrHrHrHrHrHrrr}
\hline
Par&True&\multicolumn{2}{c}{TOP}&\multicolumn{4}{c}{MJMCMC}&\multicolumn{4}{c}{RS MCMC}\\
\hline
$\Delta$&$\pi_j$&\multicolumn{2}{c}{RM}&\multicolumn{2}{c}{RM}&\multicolumn{2}{c}{MC}&\multicolumn{2}{c}{RM}&\multicolumn{2}{c}{MC}\\
\hline 
$\gamma_{4}$&0.0035&-0.0005&0.0005&-0.0019&0.0022&1.7361&2.0416&-0.0189&0.0198&1.6397&1.9768\\
$\gamma_{6}$&0.0048&-0.0006&0.0006&-0.0041&0.0051&1.8155&2.0899&-0.0241&0.0257&1.5437&1.9352\\
$\gamma_{7}$&0.0065&-0.0006&0.0006&-0.0045&0.0056&1.9763&2.3459&-0.0338&0.0353&0.2191&0.6887\\
$\gamma_{3}$&0.0076&-0.0007&0.0007&-0.0014&0.0017&2.9714&3.3660&-0.0339&0.0353&0.5167&1.2374\\
$\gamma_{8}$&0.0076&-0.0007&0.0007&-0.0066&0.0079&1.8370&2.3279&-0.0326&0.0344&1.1101&1.6163\\
$\gamma_{5}$&0.0096&-0.0007&0.0007&-0.0055&0.0075&1.5439&2.3342&-0.0430&0.0455&1.1780&1.7170\\
$\gamma_{11}$&0.0813&-0.0007&0.0007&-0.0131&0.0200&-0.7623&3.6851&-0.1060&0.1679&1.0394&2.8022\\
$\gamma_{12}$&0.0851&-0.0006&0.0006&-0.0042&0.0134&-0.4290&2.7179&-0.0637&0.0766&0.3118&1.9136\\
$\gamma_{9}$&0.1185&-0.0008&0.0008&-0.0121&0.0184&-1.3414&3.3149&-0.1277&0.1773&-0.4439&3.0463\\
$\gamma_{10}$&0.3042&-0.0006&0.0006&-0.0036&0.0071&-8.4912&9.4926&-0.0501&0.1106&2.6866&3.7344\\
$\gamma_{13}$&0.9827&-0.0002&0.0002&0.0051&0.0063&-1.6177&2.5350&0.0607&0.0638&-1.0082&1.5681\\
$\gamma_{1}$&1.0000&0.0007&0.0007&0.0000&0.0000&-4.4528&4.7091&0.0000&0.0000&-1.0018&1.2258\\
$\gamma_{2}$&1.0000&0.0000&0.0000&0.0000&0.0000&-2.3865&2.7343&0.0000&0.0000&-0.7782&0.9971\\\hline
$\text{C}(\BB\gamma)$&1.0000&1.0000&1.0000&0.9998&0.9998&0.9998&0.9998&0.9977&0.9977&0.9977&0.9977\\\hline
Eff&8192&385&385&1758&1758&1758&1758&155&155&155&155\\\hline
Tot&8192&385&385&3160&3160&3160&3160&10000&10000&10000&10000\\\hline
\end{tabular}\end{adjustbox}
\\[1pt]
\caption{\label{rmse.bias.sim.5}Average root mean squared error (RMSE) from the 100 simulated runs of MJMCMC on the epigenetic data (example 4); the values reported in the table are RMSE $\times 10^2$ for $p(\gamma_j=1|\By)$.}
\end{table}

We have addressed $p=13$ different covariates in addition to the intercept. We have considered a factor with 3 levels corresponding to whether a location belongs to a CGH, CHH or CHG genetic region, where H is either A, C or T and thus generating two covariates $X_1$ and $X_2$ corresponding to whether a location is CGH or CHH. A second factor indicates whether a distance to the previous  cytosine nucleobase (C) in DNA is 1, 2, 3, 4, 5, from 6 to 20 or greater than 20 inducing the binary covariates $X_3-X_8$. A third factor corresponds to whether a location belongs to a gene from a particular group of genes of biological interest, these groups are indicated as $M_\alpha$, $M_\gamma$, $M_\delta$ or $M_0$ inducing 3 additional covariates $X_{9}-X_{11}$.  Finally, we have considered two binary covariates $X_{12}$ and $X_{13}$ represented by expression levels exceeding 3000 and 10000, respectively. The cardinality of our search space  $\Omega$ is $2^{13} = 8192$ for this example. The correlation structure between these 13 covariates is represented in Figure~\ref{ex5corr}. 

\begin{figure}[t]
\centering
\includegraphics[width=0.48\linewidth]{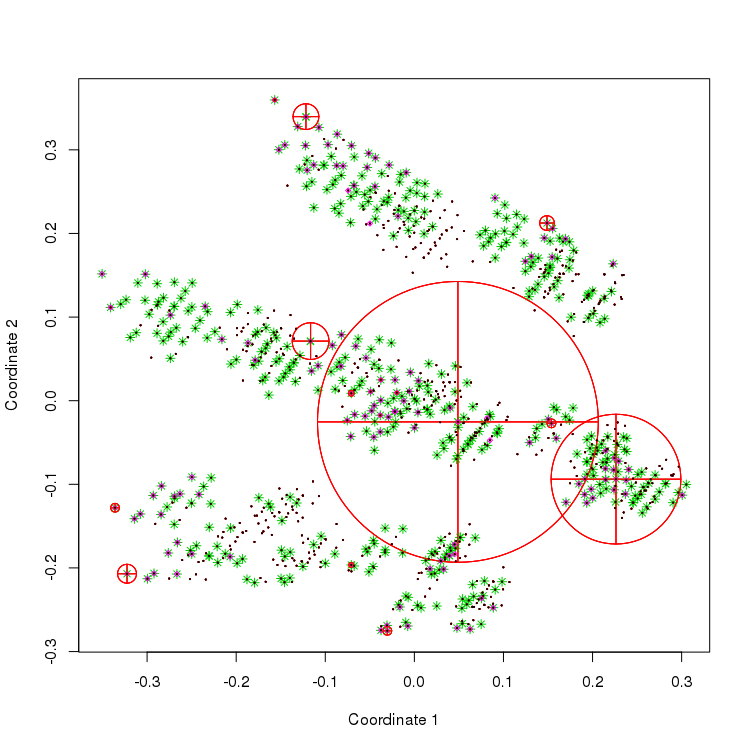}
\includegraphics[width=0.48\linewidth]{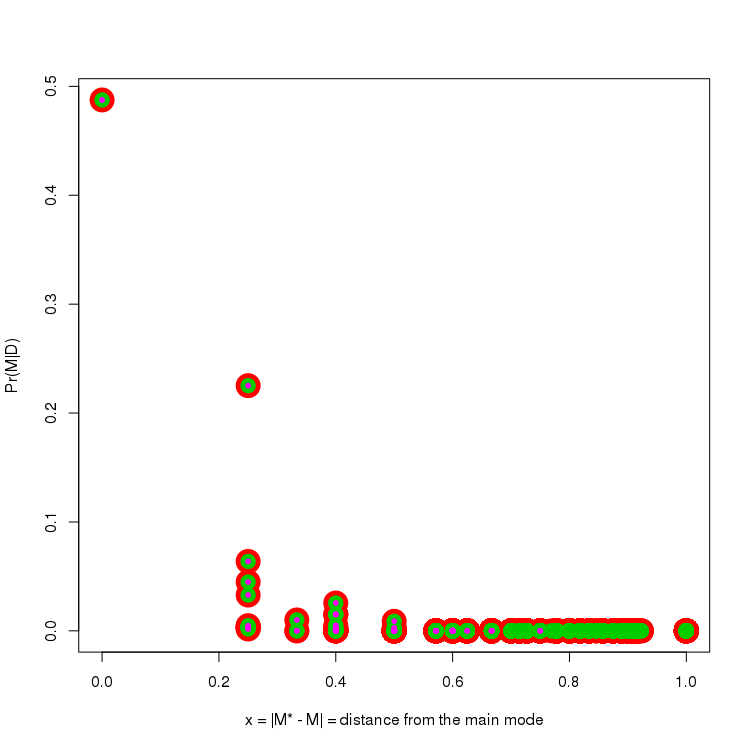}
\caption{Left: Multidimensional scale plot \citep{Rohde:2002:MBM:638950.638962} of the best 1024 models in terms of posterior model probability in the space of models (black dots are centers of the models, red circles are proportional to the posterior probabilities of models, green stars - models visited by MJMCMC, purple stars - models visited by MCMC). Right: A plot of posterior probabilities with respect to distance from the global mode (red circles correspond to all the models, the green circles - models visited by MJMCMC, the purple circles - by simple MCMC).}\label{epigpost}
\end{figure}

As seen from Table~\ref{rmse.bias.sim.5} (TOP column), within just the 385 best unique models (2.35\% of the total model space) we were able to capture  almost full posterior mass for this problem. The model space, as shown in Figure~\ref{epigpost}, has very few sparsely located modes in a quite large model space. In this example we compared MJMCMC and a simple MCMC   algorithm, the latter was allowed to only swap one component per iteration (similar to the RS algorithm within the BAS package). This example contains most of the mass in just two closely located models as can be seen in Figure \ref{epigpost}. This is why a simple RS MCMC can capture essentially most of the mass  after 10000 iterations. At the same time there are a few small modes that lie a bit further from the region of the high concentration of mass, which the simple RS MCMC algorithm did not capture. Essentially, RS MCMC stayed within a few modes for most of the time, never being able to travel to the more remote parts of the model space and generated very few (155 on average) unique models. This number is here very low compared to the total number of models visited (10000).
If there were more sparsely located remote modes, the simple RS MCMC algorithm would run into the problems similar to those discussed in the previous examples and miss a significant amount of mass. 
For MJMCMC, we ran the algorithm until 3160 models where visited, resulting in 1758 unique models. 
MJMCMC was able to capture the mass also from the remote small modes, adding a bit to the captured mass, slightly outperforming the simple RS MCMC algorithm. 
As can be seen in Table~\ref{rmse.bias.sim.5}, MJMCMC outperformed the simple RS MCMC algorithm in terms of the errors  of marginal model probabilities. Marginal inclusion probabilities in terms of RM are also more precise when MJMCMC is used. MC based approximations are also in this case worse than the RM versions, in this case with MJMCMC slightly worse.

According to marginal inclusion probabilities ($\pi_j$ column in Table~\ref{rmse.bias.sim.5}, obtained from full enumeration), factors of whether the location is CGH or CHH ($\gamma_1$ and $\gamma_2$) are both extremely significant, as well as the higher cut off for the level of expression ($\gamma_{13}$). Additionally, factors for $M_\alpha$ and $M_\delta$ groups of genes ($\gamma_9$ ad $\gamma_{10}$)  have non-zero marginal inclusion probabilities and reasonably high significance. In future it would be of interest to obtain additional covariates such as  whether a nucleobase belongs  to a particular part of the gene like the promoter or a coding region. Furthermore, it is of interest to address factors whether a base is located within a CpG island (regions with a high frequency of C bases) or whether it belongs to a transposone. Moreover, interactions of these covariates may be interesting. Alternative choices of the response distributions (e.g. binomial or negative binomial) and/or other types of random effects ($AR(k)$, $ARMA(l,k)$) might also be of an interest.

\section{Summary and discussion}\label{section5}
In this paper we have introduced the mode jumping MCMC (MJMCMC) approach for calculating posterior model probabilities and performing Bayesian model averaging and selection. The algorithm incorporates the ideas of MCMC with the possibility of large jumps combined with local optimizers to generate proposals in the discrete space of models. Unlike standard MCMC methods applied to variable selection, the developed procedure avoids getting stuck in local modes and manages to iterate through all of the important models much faster. In many cases it also outperforms Bayesian Adaptive Sampling (BAS), having the tendency to capture a higher posterior mass within the same amount of unique models visited. This can be explained by that for problems with numerous covariates BAS requires good initial marginal inclusion probabilities to perform well. 
\citet{Clyde:Ghosh:Littman:2010} demonstrated that estimates of marginal inclusion probabilities obtained from preliminary MCMC runs could largely improve BAS. A combination of MJMCMC with BAS could possibly improve both algorithms even further.

The \textit{EMJMCMC} R-package is developed and currently available from the Git Hub repository: \url{http://aliaksah.github.io/EMJMCMC2016/}. The methodology depends on the possibility of calculating marginal likelihoods within models accurately. The developed package gives a user high flexibility in the choice of methods to obtain marginal likelihoods. Whilst the default choice for marginal likelihood calculations is based on INLA \citep{rue2009eINLA}, we also have adopted efficient C based implementations for exact calculations in Bayesian linear regression and approximate calculations in Bayesian logistic and Poisson regressions in combination  with g-priors as well as other priors. Several model selection criteria for the class of methods are also addressed.  Extensive parallel computing for both MCMC moves and local optimizers is available within the developed package. Within a standard call, a user specifies how many threads are addressed within the in-build \textit{mclapply} function or \textit{snow} based parallelization. An advanced user can specify his own function to parallelize computations on both the MCMC and local optimization levels, using, for instance, modern graphical processing units - GPUs, which in turn allows additional efficiency and flexibility.

Whilst the renormalized model estimators~\eqref{approxpost} are Fisher consistent~\citep{Clyde:Ghosh:Littman:2010}, they remain generally speaking biased; although their bias reduces to zero asymptotically (with respect to the number of iterations). Standard MCMC based estimators such as~\eqref{map2}, which are both consistent and unbiased, are also available through our procedure; these estimators however tend to have a much higher variance than the aforementioned ones. As one of the further developments it would be of interest to combine knowledge available from both groups of estimators to adjust for bias and variance, which is vital for higher dimensional problems. 
 
Another aspect that requires being discussed is the model selection criterion. Different criteria  can sometimes disagree about the results of model selection. In order to avoid confusion, the researcher should be clear about the stated goals. If the goal is prediction rather than inference one should adjust for that and use AIC, WAIC~\citep{Watanabe09anintroduction} or DIC~\citep{spiegelhalter2002bayesian} rather than BIC or posterior model probability as selection criterion in MJMCMC. These choices are possible within the \textit{EMJMCMC} package as well.

Based on several experiments, we claim MJMCMC to be a rather competitive algorithm that is addressing the wide class of Generalized Linear Mixed Models (GLMM). In particular, for this class of models one can incorporate a random effect, which both can model the variability unexplained by the covariates and can introduce dependence between observations, creating additional modeling flexibility. Estimation of parameters for such models  becomes significantly harder in comparison to simple GLM. This creates the necessity to address parallel computing extensively. We have enabled the latter within our package by means of combining methods for calculating marginal likelihoods, such as the INLA methodology, and parallel MJMCMC algorithm. 

Currently, we only consider  choice of covariates to be included into the model. However, the mode jumping procedure can easily be extended to  more general cases. In the future it would be of interest to extend the procedure to  model selection and model averaging jointly across covariates, link functions, random effect structures  and response distributions. Such extensions will require even more accurate tuning of control parameters of the algorithm, introducing another important direction for further research.

\bigskip

\footnotesize

\bigskip
\begin{center}
{\large\bf ACKNOWLEDGMENTS}
\end{center}

The authors gratefully acknowledge the \textit{CELS project at the University of Oslo}, \url{http://www.mn.uio.no/math/english/research/groups/cels/index.html}, for giving the opportunity, inspiration and motivation to write this paper. The authors also thank the editor, the associate editor, and the
referees for helpful comments and suggestions which significantly improved
the manuscript. We would also like to acknowledge Paul Grini, Ole Christian Lingj{\ae}rde and Melinka Butenko  for the valuable discussions on the design of Example 4. Furthermore we want to gratitude Ole Christian Lingj{\ae}rde for the final proofreading.

\section*{References}
\bibliographystyle{agsm}
\bibliography{ref}

\newpage


\newcommand{\algorithmicbreakfor}{\textbf{break for}}
\newcommand{\algorithmicbreakwhile}{\textbf{break while}}
\renewcommand{\theequation}{A-\arabic{equation}}    
  \setcounter{equation}{0}  

\appendix

\section{Details of the MJMCMC algorithm}\label{B}

\subsection{Multiple try MCMC algorithm}
In addition to ordinary MCMC steps and mode jump MCMC, also multiple-try Metro\-polis \citep{Liu2000eMTMCMC} is considered. Multiple-try Metropolis is a sampling method that is a modified form of the Metropolis-Hastings method, designed to be able to properly parallelize the original Metropolis-Hastings algorithm. The idea of the method is to allow generating $S$ trial proposals $\Bchi^*_{1},...\Bchi^*_{S}$ in parallel from a proposal distribution $q(\cdot|\Bgamma)$. Then, $\Bgamma^* \in \{\Bchi^*_{1},...,\Bchi^*_{S}\}$ is selected with probabilities proportional to some importance weights $w(\Bgamma,\Bchi^*_i)=\pi(\Bgamma)q(\Bchi^*_i|\Bgamma)\lambda(\Bchi^*_i,\Bgamma)$ where $\lambda(\Bchi^*_i,\Bgamma)=\lambda(\Bgamma,\Bchi^*_i)$. In the reversed move  $\Bchi_{1},...\Bchi_{S-1}$ are generated from the proposal $q(\Bchi|\Bgamma^*)$ while $\Bchi_{S} = \Bgamma$. Finally, the move is accepted with probability 
\begin{equation}
r_{m}(\Bgamma,\Bgamma^*)=
{\text{min}}\left\{1,
   \frac{w(\Bchi_1^*,\Bgamma)+\ldots+w(\Bchi_S^*,\Bgamma)}
        {w(\Bchi_1,\Bgamma^*)+\ldots+w(\Bchi_S,\Bgamma^*)}
       \right\}.\label{mtmcmcacc}
 \end{equation}
 In the implementation of the algorithm, ordinary MCMC is considered as a special case of multiple try MCMC with $S=1$. 
We recommend ordinary or multiple try MCMC steps are used in at least $95\%$ of the iterations with proposals of large jumps for the remaining 5\%.

\subsection{Choice of proposal distributions}\label{app:prop}
The implementation of MJMCMC allows for great flexibility in the choices of proposal distributions for the large jumps, the local optimization and the last randomization. 
\begin{itemize}
\item Table~\ref{proposals} lists the current possibilities for drawing indexes to swap in the first large jump. One should choose distributions where a large number of components are swapped.

\item An important ingredient of the MJMCMC algorithm is the choice of
local optimizer. In the current implementation of the algorithm, several choices are possible; simulated annealing, greedy optimizers based on best neighbor optimization or first improving neighbor~\citep{Blum03metaheuristicsin} which is another variant of greedy local search accepting the first randomly selected solution better than the current.
For each alternative the neighbors are defined through swapping a few of the $\gamma_j$'s in the current model.
\item For the last randomization, again Table~\ref{proposals} lists the possibilities, but in this case a small number of swaps will be preferable. 
\end{itemize}
Different possibilities to combine the optimizers and proposals in a hybrid setting are also possible. Then, at each iteration, which proposal distributions and which optimizers to use are randomly drawn from the set of possibilities, see \citet[][sec 10.3]{robert2004bayes} for the validity of such procedures. 

\subsection{Parallel computing in local optimizers}\label{opimparal} 

General principles of utilizing multiple cores in local optimization are provided in~\citet{multcpuopt}. Given a current state $\BB\chi^*$ in the optimization routine, one can simultaneously draw several proposals $\BB\chi_1,...,\BB\chi_K$ with respect to a certain transition kernel $s_o(\cdot|\BB\gamma)$  and, if necessary, calculate the transition probabilities as the proposed models are evaluated. This step can be performed by parallel CPUs, GPUs or clusters. Consider an optimizer with the acceptance probability function $r^t_o{(\Bchi_j;\Bchi^*)}, j \in 1,...,K$, which either changes over the time (iterations) $t$ or remains unchanged. For the greedy local search $r^t_o{(\Bchi;\Bchi^*)} = \mathbb{1}\left\{\pi(\Bchi)\geq \pi(\Bchi^*)\right\}, t \in 1,2,...$. For the implemented version of the simulated annealing algorithm we consider $r^t_o{(\Bchi;\Bchi^*)} = \min\left\{1,\exp\left({\frac{\log\pi(\Bchi)-\log\pi(\Bchi^*)}{T_t}}\right)\right\}, i \in 1,...,N$, where $T_t$ is the SA temperature~\citep{Blum03metaheuristicsin} parameter at iteration $t$. The proposed parallelization strategy is given in detail in Algorithm~\ref{paralopt}.
\begin{algorithm}[h]
\caption{Parallel optimization}\label{paralopt}
\begin{algorithmic}[1]
\Procedure{Optimize}{N}
\State $\Bchi^* \gets \Bchi^*_{0}$
\For{$i=1,...,N$}
\State $\Bchi_{i,1},...,\Bchi_{i,K} \sim s_o(\cdot|\Bchi^*)$ \Comment{make $K$ proposals in parallel}
\State\Comment{and calculate marginal likelihoods}
\For{$j=1,...,K$}
\State $r \gets r^i_o{(\Bchi_{i,j};\Bchi^*)}$ \Comment{calculate acceptance probability}
\If{$ \text{Unif}[0;1] \leq r$} 
\State $\Bchi^* \gets \Bchi_{i,j}$ \Comment{accept the transition}
\EndIf
\EndFor
\State $\Bchi^*_{i} \gets \Bchi^*$ 
\EndFor
\State\textbf{return} $\Bchi^*_{N}$
\EndProcedure
\end{algorithmic}
\end{algorithm}

\subsection{Parallel MJMCMC with a mixture of proposals}\label{mjmcmcparal} 

Here we described the full version of our algorithm based on a combination of  Algorithm~\ref{MJMCMCalg01} and the multiple try idea. The suggested  MJMCMC approach allows to both jump between local modes efficiently and to explore the solutions around the modes simultaneously  whilst keeping the desired ergodicity of the MJMCMC procedure. 
This implementation allows for mixtures of both local optimizers and proposals to be addressed within MJMCMC. 
Both the local optimization and the multiple try steps utilize multiple CPUs and GPUs of a single machine or a cluster of nodes.
The pseudo-code of the algorithm is given in Algorithm \ref{MJMCMCalg} below. In this pseudo-code we consider the following notation:
\begin{itemize}
\item $\varrho$ - the probability for a large jump;
\item $\mathsf{P}_\mathsf{o}(\cdot)$ - the distribution for the choice of the local optimizers, a discrete distribution over a finite number of possibilities; 
\item $\mathsf{P}_\mathsf{l}(\cdot)$ - the distribution for the choice of large jump transition kernel, a discrete distribution over the possibilities in Table~\ref{proposals} with high probabilities on a large number of swaps; 
\item $\mathsf{P}_\mathsf{r}(\cdot)$ - the distribution for the choice of the randomizing kernel, a discrete distribution over a finite number of possibilities, also from Table~\ref{proposals}, but with a small number of changes; 
\item $\mathsf{P}_\mathsf{g}(\cdot)$ - the distribution for the choice of proposals within the multiple try MCMC, a discrete distribution over the possibilities in Table~\ref{proposals} with a high probability on a small number of swaps.
\end{itemize}
\begin{algorithm}[h]
\caption{Mode jumping MCMC}\label{MJMCMCalg}
\begin{algorithmic}[1]
\Procedure{MJMCMC}{\textit{Numit}}
	\State $\Bgamma\gets \Bgamma_0$	\Comment{define the initial state}
	\For{$t=1,...,Numit$}
	\If{$\text{Unif}[0;1]\leq\varrho$}\Comment{large jump with local optimization}
	\State $q_{l}\sim \mathsf{P}_\mathsf{l}(\boldsymbol{\cdot})$ \Comment{choose large jump kernel}	
	\State $q_{o} \sim \mathsf{P}_\mathsf{o}(\boldsymbol{\cdot})$ \Comment{choose local optimizer}
	\State $q_{r}\sim \mathsf{P}_\mathsf{r}(\boldsymbol{\cdot})$ \Comment{choose randomization kernel}
	\State $I\sim q_l(\cdot|\Bgamma)$\Comment{Indices for large jump}
	\State  $\Bchi^*_0 \gets \text{SWAP}(\Bgamma,I)$\Comment{large jump}
	\State $\Bchi^*_\mathsf{k}\sim q_{o}(\cdot|\Bchi^*_0)$ \Comment{local optimization} 
\State $\Bgamma^*\sim q_{r}(\cdot|\Bchi^*_\mathsf{k})$ \Comment{randomization around the mode}
\State $\Bchi_0\gets \text{SWAP}(\Bgamma^*,I)$ \Comment{reverse large jump}
\State $\Bchi_\mathsf{k}\sim q_{o}(\cdot|\Bchi_0)$ \Comment{local optimization} 
\State $r \gets r_m{(\Bchi,\Bgamma;\Bchi^*,\Bgamma^*)}$ \Comment{from~\eqref{locmcmcgen01}}
\Else \Comment{ordinary proposal}
\State $q_{g}\sim \mathsf{P}_\mathsf{g}(\boldsymbol{\cdot})$ \Comment{choose multiple try proposal kernel}
\State $\Bgamma^*\sim q_{g}(\cdot|\Bgamma)$\Comment{proposed solution}
\State $r \gets r_m(\Bgamma,\Bgamma^*)$\Comment{from~\eqref{mtmcmcacc}}
\EndIf
\If{$ \text{Unif}[0;1] \leq r$} 
\State $\Bgamma\gets \Bgamma^*$ \Comment{accept the move}
\EndIf
\EndFor 
\EndProcedure
\end{algorithmic}
\end{algorithm}
The essential ingredients  of the parallel version of the MJMCMC with a mixture of proposals (Algorithm~\ref{MJMCMCalg}) are as follows:
\begin{itemize}
\item Multiple try MCMC steps are performed for the steps with no mode jumps;
\item At the iterations with mode jumps the large jump proposals $q_{l}\sim \mathsf{P}_\mathsf{l}(\boldsymbol{\zeta})$, the optimization proposals $q_{o}\sim \mathsf{P}_\mathsf{o}(\boldsymbol{\zeta})$, and  the randomizing kernels $q_{r}\sim \mathsf{P}_\mathsf{r}(\boldsymbol{\zeta})$ are chosen randomly;
\item At the iterations with no mode jumps the proposal is chosen randomly as  $q_g \sim \mathsf{P}_\mathsf{g}(\boldsymbol{\zeta})$;
\item The optimization steps are parallelized as described in~\ref{opimparal}.
\item The multiple-try steps are parallelized.
\end{itemize}

\section{Supplementary materials for the experiments}\label{appccc}
\setcounter{table}{0}
Table~\ref{optimparam} describes some of the tuning parameters used for the different examples. Here, MTMCMC refers to the multiple try MCMC steps. The remaining tuning parameters, describing the mixture distributions $P_0, P_l$ and $P_r$ are specified in tables~\ref{proposals1} (example 1), \ref{proposals3} (example 2),  \ref{proposals4} (example 3) and
\ref{proposals5} (example 4).  

\begin{table}[H]
\begin{adjustbox}{center}
\begin{tabular}{ 
|c|c|c|c|c|c|c|c|c|c|c|c|c|c|}
\hline
Example&CPU&\multicolumn{4}{|c|}{SA}&\multicolumn{3}{|c|}{Greedy}&\multicolumn{2}{|c|}{MT}\\\hline
No&Num&$S_t$&$\Delta t$&$t_0$&$t_f$&S&LS&FI&Size&Steps\\\hline
1&4&4&3&10&14$\times 10^{-5}$&15&F&T&4&15\\\hline
2&2&5&3&10&14$\times 10^{-5}$&20&F&T&2&20\\\hline
3&10&18&3&10&14$\times 10^{-5}$&88&F&T&10&88\\\hline
4&1&3&3&10&14$\times 10^{-5}$&13&F&T&2&13\\\hline
S.1&4&4&3&10&14$\times 10^{-5}$&15&F&T&4&15\\\hline
\end{tabular}
\end{adjustbox}
\\[1pt]
\caption{\label{optimparam}Tuning parameters for local optimization within MJMCMC in the examples (Example No); CPU (Num) - the number of CPUs utilized within the examples; $S_t$ - number of iterations per temperature in SA algorithm; $\Delta t$ - cooling factor of the cooling schedule of SA algorithm; $t_0$ - initial temperature of SA algorithm; $t_f$ - final temperature of SA algorithm; S - number of iterations in Greedy algorithm (per run); LS - if local stop is allowed in Greedy algorithm; FI - if the first improving neighbor strategy is applied in Greedy algorithm; Size - number of proposals per step in the multiple try steps; Steps - number of multiple try iterations within the local optimizer.}
\end{table}

\begin{table}[H]
\resizebox{\textwidth}{!}{%
\begin{tabular}{ 
|c|c|c|c|c|c|Hc|c|c|c|}
\hline
Proposal&Optimizer&Frequency&Type 1&Type 4&Type 3&Type x&Type 5&Type 6&Type 2\\\hline
$q_{g}$&-&$\varrho=0.9836$&0.1176&0.3348&0.2772&0.0000&0.0199&0.2453&0.0042\\\hline
$S$&-&-&$\{2,2\}$&2&$\{2,2\}$&$2$&$1$&$1$&$15$\\\hline
$\rho_j$&-&-&$\widehat{p}(\gamma_j|\By)$&-&-&-&-&-&$\widehat{p}(\gamma_j|\By)$\\\hline
\hline
$q_{l}$&-&0.0164&0&1&0&0&0&0&0\\\hline
$S$&-&-&-&4&$-$&$-$&$-$&$-$&$-$\\\hline
$\rho_i$&-&-&-&-&-&-&-&-&-\\\hline\hline
$q_{o}$&SA&0.5553&0.0788&0.3942&0.1908&0.0000&0.1928&0.1385&0.0040\\\hline
$q_{o}$&GREEDY&0.2404&0.0190&0.3661&0.2111&0.0000&0.2935&0.1046&0.0044\\\hline
$q_{o}$&MTMCMC&0.2043&0.2866&0.1305&0.2329&0.0000&0.1369&0.2087&0.0040\\\hline
$S$&-&-&$\{2,2\}$&2&$\{2,2\}$&$2$&$1$&$1$&$15$\\\hline
$\rho_j$&-&-&$\widehat{p}(\gamma_j|\By)$&-&-&-&-&-&$\widehat{p}(\gamma_j|\By)$\\\hline
\hline
$q_{r}$&-&-&0&0&0&0&0&0&$1$\\\hline
$S$&-&-&-&-&$-$&$-$&$-$&$-$&$15$\\\hline
$\rho_j$&-&-&-&-&-&-&-&-&$0.0010$\\\hline
\end{tabular}
}
\caption{Other tuning parameters of MJMCMC for all proposal types ($q_g, g_l, q_o$, and $q_r$) in example 1; Optimizer - to which optimizer the proposal belongs (if not relevant "-"); Frequency - the frequency at which the proposal is addressed ($\varrho$ for $q_g$ and $1-\varrho$ for $q_l$) and the frequency within the set of local optimizers ($\mathsf{P}_\mathsf{o}$ for local optimizers); Type X - the frequency of proposal of type X Table~\ref{proposals}; $S$ - maximal allowed size of the neighborhood for the corresponding proposal; $\rho_i$ - probability of change of component $i$ of the current solution (if applicable to the proposal), where $\widehat{p}(\gamma_j|\By) = \widehat{p}(\gamma_j=1|\By)$ are the approximations of marginal inclusion probabilities. Notice that for MJMCMC* reported in the example only proposals of type 4 are used. }\label{proposals1}
\end{table}

\begin{table}[H]
\resizebox{\textwidth}{!}{%
\begin{tabular}{|c|c|c|c|c|c|Hc|c|c|c|}
\hline
Proposal&Optimizer&Frequency&Type 1&Type 4&Type 3&Type x&Type 5&Type 6&Type 2\\\hline
$q_{g}$&-&$\varrho=0.9820$&0.1179&0.3357&0.2779&0.0000&0.0200&0.2459&0.0021\\\hline
$S$&-&-&$\{1,1\}$&1&$\{1,1\}$&$1$&$1$&$1$&$20$\\\hline
$\rho_j$&-&-&$\widehat{p}(\gamma_j|\By)$&-&-&-&-&-&$\widehat{p}(\gamma_j|\By)$\\\hline
\hline
$q_{l}$&-&0.0180&0&1&0&0&0&0&0\\\hline
$S$&-&-&-&5&$-$&$-$&$-$&$-$&$-$\\\hline
$\rho_i$&-&-&-&-&-&-&-&-&-\\\hline\hline
$q_{o}$&SA&0.5042&0.0636&0.3249&0.1571&0.0000&0.2288&0.2246&0.0009\\\hline
$q_{o}$&GREEDY&0.2183&0.0160&0.3085&0.1779&0.0000&0.2474&0.2493&0.0007\\\hline
$q_{o}$&MTMCMC&0.2774&0.2879&0.3016&0.1582&0.0000&0.1107&0.1401&0.0013\\\hline
$S$&-&-&$\{1,1\}$&1&$\{1,1\}$&$1$&$1$&$1$&$20$\\\hline
$\rho_j$&-&-&$\widehat{p}(\gamma_j|\By)$&-&-&-&-&-&$\widehat{p}(\gamma_j|\By)$\\\hline
\hline
$q_{r}$&-&-&0&0&0&0&0&0&$1$\\\hline
$S$&-&-&-&-&$-$&$-$&$-$&$-$&$20$\\\hline
$\rho_j$&-&-&-&-&-&-&-&-&$0.0010$\\\hline
\end{tabular}
}
\caption{Other tuning parameters of MJMCMC for all proposal types ($q_g, g_l, q_o$, and $q_r$) in example 2; see Tables~\ref{proposals} and~\ref{proposals1}  for details.}\label{proposals3}
\end{table}

\begin{table}[H]
\resizebox{\textwidth}{!}{%
\begin{tabular}{|c|c|c|c|c|c|Hc|c|c|c|}
\hline
Proposal&Optimizer&Frequency&Type 1&Type 4&Type 3&Type x&Type 5&Type 6&Type 2\\\hline
$q_{g}$&-&$\varrho=0.9816$&0.0932&0.2654&0.2197&-&0.0158&0.1944&0.2116\\\hline
$S$&-&-&$\{1,3\}$&3&$\{1,3\}$&$3$&$1$&$1$&$88$\\\hline
$\rho_j$&-&-&$\widehat{p}(\gamma_j|\By)$&-&-&-&-&-&$\widehat{p}(\gamma_j|\By)$\\\hline
\hline
$q_{l}$&-&0.0164&0&1&0&0&0&0&0\\\hline
$S$&-&-&-&20&$-$&$-$&$-$&$-$&$-$\\\hline
$\rho_i$&-&-&-&-&-&-&-&-&-\\\hline\hline
$q_{o}$&SA&0.5553&0.0633&0.3165&0.1532&-&0.1548&0.1112&0.2011\\\hline
$q_{o}$&GREEDY&0.2404&0.0149&0.2871&0.1656&-&0.2302&0.0820&0.2201\\\hline
$q_{o}$&MTMCMC&0.2043&0.2310&0.1052&0.1877&-&0.1103&0.1682&0.1980\\\hline
$S$&-&-&$\{1,3\}$&3&$\{1,3\}$&$3$&$1$&$1$&$88$\\\hline
$\rho_j$&-&-&$\widehat{p}(\gamma_j|\By)$&-&-&-&-&-&$\widehat{p}(\gamma_j|\By)$\\\hline
\hline
$q_{r}$&-&-&0&0&0&0&0&0&$1$\\\hline
$S$&-&-&-&-&$-$&$-$&$-$&$-$&$88$\\\hline
$\rho_j$&-&-&-&-&-&-&-&-&$0.0010$\\\hline
\end{tabular}
}
\caption{Other tuning parameters of MJMCMC for all proposal types ($q_g, g_l, q_o$, and $q_r$) in example 3; see Table~\ref{proposals} and~\ref{proposals1}  for details.}\label{proposals4}
\end{table}

\begin{table}[H]
\resizebox{\textwidth}{!}{%
\begin{tabular}{|c|c|c|c|c|c|Hc|c|c|c|}
\hline
Proposal&Optimizer&Frequency&Type 1&Type 4&Type 3&Type x&Type 5&Type 6&Type 2\\\hline
$q_{g}$&-&$\varrho=0.9615$&0.1662&0.3323&0.1662&0.0000&0.1662&0.1662&0.0029\\\hline
$S$&-&-&$\{1,1\}$&1&$\{1,1\}$&$1$&$1$&$1$&$13$\\\hline
$\rho_j$&-&-&$\widehat{p}(\gamma_j|\By)$&-&-&-&-&-&$\widehat{p}(\gamma_j|\By)$\\\hline
\hline
$q_{l}$&-&0.0385&0&1&0&0&0&0&0\\\hline
$S$&-&-&-&4&$-$&$-$&$-$&$-$&$-$\\\hline
$\rho_i$&-&-&-&-&-&-&-&-&-\\\hline\hline
$q_{o}$&SA&0.5000&0.0657&0.3281&0.1588&0.0000&0.2247&0.2209&0.0019\\\hline
$q_{o}$&GREEDY&0.2500&0.0160&0.3083&0.1778&0.0000&0.2472&0.2491&0.0014\\\hline
$q_{o}$&MTMCMC&0.2500&0.2875&0.3012&0.1580&0.0000&0.1105&0.1398&0.0026\\\hline
$S$&-&-&$\{1,1\}$&1&$\{1,1\}$&$1$&$1$&$1$&$13$\\\hline
$\rho_j$&-&-&$\widehat{p}(\gamma_j|\By)$&-&-&-&-&-&$\widehat{p}(\gamma_j|\By)$\\\hline
\hline
$q_{r}$&-&-&0&0&0&0&0&0&$1$\\\hline
$S$&-&-&-&-&$-$&$-$&$-$&$-$&$13$\\\hline
$\rho_j$&-&-&-&-&-&-&-&-&$0.0010$\\\hline
\end{tabular}
}
\caption{Other tuning parameters of MTMCMC for all proposal types ($q_g, g_l, q_o$, and $q_r$) in example 4; see Table~\ref{proposals1} and~\ref{proposals}  for details.}\label{proposals5}
\end{table}

\subsection{Details on example 2}\label{C4}

In the addressed data set the true regression parameters were chosen to be $\beta_0 = 99$ for the intercept, and for the slope coefficients 
\[
\boldsymbol{\beta} =(-4, 0, 1, 0, 0, 0, 1, 0, 0, 0, 1.2, 0, 37.1, 0, 0, 50, -0.00005, 10, 3, 0).
\]
What concerns the covariates, $X_1$ and $X_3$ are factors from a group with 3 levels, $X_4$ and $X_6$ are  from another group with 3 levels but additionally  correlated with $X_1$ and $X_3$, $X_7$ and $X_8$ are two exponentially distributed variables with rate 0.3 jointly made dependent through copulas, $X_9, X_{10}$ and $X_{11}$ are all uniformly distributed with range from -1 to 10 and also jointly dependent through copulas, $X_{12}, X_{13}, X_{14}$ and $X_{15}$ are multivariate normal with a zero mean, standard deviation of 0.2 and some covariance structure,   $X_{16}$ represents some seasonality incorporated by the sinus transformation of the radiant representation of some angle equal to the corresponding ordering numbers of observations, $X_{17}$ is the quadratic trend associated to the squared value of positions of observations, $X_{19} = (-4 + 5X_1 + 6X_3)X_{15}$ and $X_{20} = (-4 + 5X_1 + 6X_3)X_{11}$, finally to avoid over specification 2 layers from the mentioned above groups of factors were replaced with some auxiliary covariates $X_2 = (X_{10}+X_{14})\times X_9$ and $X_5 = (X_{11}+X_{15})\times X_{12}$. The linear predictor is drawn as $\eta \sim N(\beta'X, 0.5)$, whilst the observations $Y$ are independent Bernoulli variables with the probability of success modeled by a logit transformation of the linear predictor, namely $Y\sim Bernoulli\left(p = \frac{\exp(\eta)}{1+\exp(\eta)}\right)$.

\section{Further results}\label{sec:further.results}
\setcounter{table}{0}
In tables~\ref{bias.sim.2} (example 1), \ref{bias.sim.3} (example 2) and \ref{bias.sim.5} (example 4) the
estimated biases, corresponding to the RMSE estimates given in tables \ref{rmse.sim.2}, \ref{rmse.sim.3} 
 and \ref{rmse.bias.sim.5}, are reported. In addition, an extra simulation experiment on linear regression based on simulated data is reported in~\ref{sec:example.s.1}.

\begin{table}[H]
\resizebox{\textwidth}{!}{%
\begin{tabular}{crrrrrrrHrrrrHrHr}
\hline
Par&True&TOP&\multicolumn{2}{c}{MJMCMC}&\multicolumn{2}{c}{MJMCMC$^2$}&\multicolumn{1}{c}{BAS}&&\multicolumn{2}{c}{$\text{MC}^3$}&\multicolumn{2}{c}{RS}&\multicolumn{4}{c}{MJMCMC*}\\
\hline 
$\Delta$&$\pi_j$&-&RM&MC&RM&MC&RM&RM&MC&RM&MC&RM&RM&RM&MC&MC\\\hline
$\gamma_{8}$&0.16&-3.51&-6.54&-10.28&-5.09&-9.64&-5.19&-6.31&5.37&-3.20&4.96&-3.06&-6.19&6.23&-7.74&9.06\\
$\gamma_{13}$&0.16&-3.34&-7.44&-10.12&-5.57&-9.94&-6.25&-7.17&7.46&2.86&8.06&2.65&-6.31&6.38&-5.45&10.54\\
$\gamma_{14}$&0.19&-3.24&-8.27&-11.69&-6.28&-11.93&-6.19&-7.36&5.27&-1.86&5.37&-2.03&-7.10&7.15&-7.77&10.91\\
$\gamma_{12}$&0.22&-3.27&-6.82&-12.91&-5.54&-13.15&-3.08&-5.14&3.00&-5.82&3.76&-5.06&-5.14&5.29&-3.30&10.93\\
$\gamma_{5}$&0.23&-2.56&-6.21&-12.71&-4.55&-13.35&-1.80&-3.88&-4.79&-12.98&-4.28&-12.72&-5.22&5.39&-9.01&10.90\\
$\gamma_{9}$&0.23&-3.27&-9.45&-15.67&-7.35&-16.11&-9.26&-8.31&4.53&-2.45&4.33&-2.10&-7.58&7.68&-8.54&11.06\\
$\gamma_{7}$&0.29&-2.31&-4.15&-12.04&-3.41&-12.36&-2.24&-2.85&-0.47&-9.41&-1.00&-9.56&-3.64&3.91&-5.81&10.10\\
$\gamma_{4}$&0.30&-1.57&-5.82&-18.74&-3.67&-17.10&0.85&-0.67&-12.67&-21.79&-13.24&-21.45&-4.12&4.63&-9.60&13.22\\
$\gamma_{6}$&0.33&-1.92&-8.49&-19.07&-6.09&-18.84&-3.06&-5.21&8.99&7.16&10.09&6.81&-5.59&5.87&-12.00&15.43\\
$\gamma_{1}$&0.34&-2.51&-11.25&-21.94&-7.25&-20.29&-8.42&-7.13&22.36&25.10&23.32&24.63&-7.35&7.58&-9.15&12.97\\
$\gamma_{3}$&0.39&-0.43&3.51&-7.20&2.09&-4.43&4.98&3.51&-21.11&-30.20&-21.13&-29.92&1.86&2.99&-7.62&12.66\\
$\gamma_{2}$&0.57&1.58&5.66&-8.73&3.71&-7.51&13.73&8.38&-30.41&-37.52&-29.05&-37.12&4.57&5.11&-10.22&14.04\\
$\gamma_{11}$&0.59&0.58&2.86&11.75&2.13&15.32&-3.95&-1.14&10.67&21.68&10.29&21.23&1.61&2.77&-5.58&12.77\\
$\gamma_{10}$&0.77&3.25&7.50&-2.57&5.91&2.33&15.42&10.33&-21.22&-19.06&-20.01&-19.55&6.25&6.41&-10.59&14.27\\
$\gamma_{15}$&0.82&3.48&9.17&0.22&6.85&3.65&14.50&11.64&-69.61&-76.81&-69.14&-76.30&6.59&6.75&-10.72&14.76\\\hline
$\text{C}(\Bgamma)$&1.00&0.86&0.58&0.58&0.71&0.71&0.66&0.67&0.10&0.10&0.10&0.10&0.60&0.60&0.60&0.60\\\hline
Eff&$2^{15}$&3276&1909&1909&3237&3237&3276&3276&829&829&1071&1071&3264&3264&3264&3264\\\hline
Tot&$2^{15}$&3276&3276&3276&5936&5936&3276&3276&3276&3276&3276&3276&4295&4295&4295&4295\\\hline
\end{tabular}
}
\\[1pt]
\caption{\label{bias.sim.2}Bias for the 100 simulated runs of every algorithm on the Crime data (example 1); the values reported in
the table are Bias $\times 10^2$ for $p(\gamma_j=1|\By)$. See the caption of Table~\ref{rmse.sim.2} for further details.}
\end{table}

\begin{table}[H]
\resizebox{\textwidth}{!}{%
\begin{tabular}{cccccccccccc}
\hline
Par&True&TOP&\multicolumn{4}{c}{MJMCMC}&BAS&BAS-RS&\multicolumn{2}{c}{RS}\\
\hline 
$\Delta$&$\pi_j$&-&RM&MC&RM&MC&RM&RM&RM&MC\\ \hline
 $\gamma_{6}$&0.29&0.00&-7.23&-14.89&-4.48&-16.40&-6.46&-3.59&-5.96&0.23\\ 
 $\gamma_{8}$&0.31&0.00&-5.97&-13.94&-3.89&-16.57&-5.57&-2.85&-5.28&-0.35\\ 
 $\gamma_{12}$&0.35&0.00&-4.07&-8.12&-2.56&-11.65&-4.20&-1.82&-3.80&0.06\\ 
 $\gamma_{15}$&0.35&0.00&-3.66&-8.85&-2.21&-12.04&-4.58&-1.35&-3.25&-0.28\\ 
 $\gamma_{2}$&0.36&0.00&-4.60&-14.71&-2.81&-16.80&-5.39&-2.19&-3.51&0.04\\ 
 $\gamma_{20}$&0.37&0.00&-4.16&-8.38&-2.46&-12.03&-3.30&-1.75&-4.07&-0.12\\ 
 $\gamma_{3}$&0.40&0.00&-8.99&-19.22&-5.58&-21.72&-9.73&-4.63&-6.69&0.23\\ 
 $\gamma_{14}$&0.44&0.00&1.08&7.12&0.51&7.63&3.68&-0.62&-0.99&0.22\\ 
 $\gamma_{10}$&0.44&0.00&-2.68&-7.62&-1.68&-11.89&-4.79&-0.29&-1.19&0.13\\ 
 $\gamma_{5}$&0.46&0.00&-1.74&-10.78&-0.88&-12.29&-3.93&0.57&0.55&-0.23\\ 
 $\gamma_{9}$&0.61&0.00&0.32&-2.29&0.00&-1.24&3.78&0.22&1.99&-0.11\\ 
 $\gamma_{4}$&0.88&0.00&5.61&6.20&3.71&6.13&6.60&5.54&7.58&-0.45\\ 
 $\gamma_{11}$&0.91&0.00&5.36&6.47&3.87&6.84&4.64&3.01&4.29&-0.28\\ 
 $\gamma_{1}$&0.97&0.00&1.86&0.98&1.32&1.17&2.43&1.94&2.28&-0.31\\ 
 $\gamma_{13}$&1.00&0.00&0.00&-0.33&0.00&-0.29&0.00&0.00&0.00&-0.3\\ 
 $\gamma_{7}$&1.00&0.00&0.00&-0.41&0.00&-0.36&0.00&0.00&0.00&-0.27\\ 
 $\gamma_{16}$&1.00&0.00&0.00&-0.33&0.00&-0.31&0.00&0.00&0.00&-0.17\\ 
 $\gamma_{17}$&1.00&0.00&0.00&-0.38&0.00&-0.35&0.00&0.00&0.00&-0.17\\ 
 $\gamma_{18}$&1.00&0.00&0.00&-0.37&0.00&-0.32&0.00&0.00&0.00&-0.19\\ 
 $\gamma_{19}$&1.00&0.00&0.00&-0.40&0.00&-0.32&0.00&0.00&0.00&-0.34\\ 
\hline $\text{C}(\BB\gamma)$&1.00&1.00&0.72&0.72&0.85&0.85&0.74&0.85&0.68&0.68\\ 
\hline Eff&$2^{20}$&10000&5148&5148&9988&9988&10000&10000&1889&1889\\ 
\hline Tot&$2^{20}$&10000&9998&9998&19849&19849&10000&10000&10000&10000\\ 
\hline 
\end{tabular}
}
\\[1pt]
\caption{\label{bias.sim.3}Bias for the 100 simulated runs of every algorithm on the simulated data of experiment 2; the values reported in
the table are Bias $\times 10^2$ for $p(\gamma_j=1|\By)$.} See the caption of Table~\ref{rmse.sim.2} for further details.
\end{table}

\begin{table}[!t]
\centering
\resizebox{0.8\textwidth}{!}{%
\begin{tabular}{crrHrHrHrHrHrHrH}
\hline
Par&True&\multicolumn{2}{c}{TOP}&\multicolumn{4}{c}{MJMCMC}&\multicolumn{4}{c}{RS}\\
\hline
$\Delta$&$\pi_j$&\multicolumn{2}{c}{RM}&\multicolumn{2}{c}{RM}&\multicolumn{2}{c}{MC}&\multicolumn{2}{c}{RM}&\multicolumn{2}{c}{MC}\\
\hline 
$\gamma_{4}$&0.0035&-0.0005&0.0005&-0.0019&0.0022&1.7361&2.0416&-0.0189&0.0198&1.6397&1.9768\\
$\gamma_{6}$&0.0048&-0.0006&0.0006&-0.0041&0.0051&1.8155&2.0899&-0.0241&0.0257&1.5437&1.9352\\
$\gamma_{7}$&0.0065&-0.0006&0.0006&-0.0045&0.0056&1.9763&2.3459&-0.0338&0.0353&0.2191&0.6887\\
$\gamma_{3}$&0.0076&-0.0007&0.0007&-0.0014&0.0017&2.9714&3.3660&-0.0339&0.0353&0.5167&1.2374\\
$\gamma_{8}$&0.0076&-0.0007&0.0007&-0.0066&0.0079&1.8370&2.3279&-0.0326&0.0344&1.1101&1.6163\\
$\gamma_{5}$&0.0096&-0.0007&0.0007&-0.0055&0.0075&1.5439&2.3342&-0.0430&0.0455&1.1780&1.7170\\
$\gamma_{11}$&0.0813&-0.0007&0.0007&-0.0131&0.0200&-0.7623&3.6851&-0.1060&0.1679&1.0394&2.8022\\
$\gamma_{12}$&0.0851&-0.0006&0.0006&-0.0042&0.0134&-0.4290&2.7179&-0.0637&0.0766&0.3118&1.9136\\
$\gamma_{9}$&0.1185&-0.0008&0.0008&-0.0121&0.0184&-1.3414&3.3149&-0.1277&0.1773&-0.4439&3.0463\\
$\gamma_{10}$&0.3042&-0.0006&0.0006&-0.0036&0.0071&-8.4912&9.4926&-0.0501&0.1106&2.6866&3.7344\\
$\gamma_{13}$&0.9827&-0.0002&0.0002&0.0051&0.0063&-1.6177&2.5350&0.0607&0.0638&-1.0082&1.5681\\
$\gamma_{1}$&1.0000&0.0007&0.0007&0.0000&0.0000&-4.4528&4.7091&0.0000&0.0000&-1.0018&1.2258\\
$\gamma_{2}$&1.0000&0.0000&0.0000&0.0000&0.0000&-2.3865&2.7343&0.0000&0.0000&-0.7782&0.9971\\\hline
$\text{C}(\BB\gamma)$&1.0000&1.0000&1.0000&0.9998&0.9998&0.9998&0.9998&0.9977&0.9977&0.9977&0.9977\\\hline
Eff&8192&385&385&1758&1758&1758&1758&155&155&155&155\\\hline
Tot&8192&385&385&3160&3160&3160&3160&10000&10000&10000&10000\\\hline
\end{tabular}
}
\\[1pt]
\caption{\label{bias.sim.5}Bias of the mean squared error (BIAS) from the 100 simulated runs of MJMCMC on the epigenetic data (example 4); the values reported in
the table are BIAS $\times 10^2$ for $p(\gamma_j=1|\By)$. See the caption of Table~\ref{rmse.sim.2} for further details.}
\end{table}

\subsection{Example S.1}\label{sec:example.s.1}

In this experiment we compared MJMCMC to BAS and competing MCMC methods ($\text{MC}^3$, RS) using simulated data following the same linear Gaussian regression model as~\citet{Clyde:Ghosh:Littman:2010} with $p = 15$ and $n = 100$.  All columns of the design matrix except for the ninth were generated from independent standard normal random variables and then centered. The ninth column was constructed so that its correlation with the second column was approximately 0.99. The regression parameters were chosen as $\beta_0 = 2$,  $\boldsymbol{\beta} =(-0.48,8.72,-1.76,-1.87,0,0,0,0,4,0,0,0,0,0,0)$ while the variance used was $\sigma^2 = 1$. 

When performing inference, Zellner's g-prior with
$g = T$ was used  for the regression parameters within each model. The marginal likelihood of a model could then be calculated through~\eqref{gelmlik}.
To complete the prior specification, we used~\eqref{glmgammaprior} with $q = 0.5$. This lead to a rather simple example with two main modes in the model space. Simple approaches were expected to work well in this case.  
The exact posterior model probabilities could be obtained by enumeration of the model space in this case, making comparison with the truth possible. 

In the BAS algorithm 3276 models unique were visited (about 10\% of the total number of models). When running the MCMC algorithms approximately the same number of iterations were used. For the MJMCMC algorithm, calculation of marginal likelihoods of models were stored making it unnecessary to recompute these when a model was revisited. Therefore,  for MJMCMC also a number of iterations giving the number of \emph{unique} models visited comparable with BAS was included.
For each algorithm  100 replications were performed. 
\begin{table}[!t]
\resizebox{\textwidth}{!}{%
\begin{tabular}{crrrrrrrHrrrr}
\hline
Par&True&TOP&\multicolumn{2}{c}{MJMCMC}&\multicolumn{2}{c}{MJMCMC$^2$}&\multicolumn{1}{c}{BAS}&&\multicolumn{2}{c}{$\text{MC}^3$}&\multicolumn{2}{c}{RS}\\
\hline 
$\Delta$&$\pi_j$&-&RM&MC&RM&MC&RM&unif&MC&RM&MC&RM\\ \hline
$\gamma_{12}$&0.09&0.29&2.11&5.31&1.19&5.73&1.23&1.35&2.77&4.27&2.14&3.83\\
$\gamma_{14}$&0.10&0.28&2.13&6.99&1.13&6.25&1.14&1.26&2.92&4.31&2.59&3.95\\
$\gamma_{10}$&0.11&0.28&2.31&7.41&1.31&7.74&1.15&1.31&3.06&4.31&2.40&4.07\\
$\gamma_8$&0.12&0.27&1.97&6.44&1.09&7.80&0.97&1.12&2.77&4.01&2.23&3.87\\
$\gamma_6$&0.13&0.25&2.25&8.87&1.27&8.46&1.05&1.28&3.12&4.74&2.72&4.31\\
$\gamma_7$&0.14&0.25&2.06&7.75&1.29&8.51&1.05&1.19&3.45&4.52&2.50&4.17\\
$\gamma_{13}$&0.15&0.24&2.42&9.98&1.36&8.79&1.15&1.24&3.50&4.87&2.44&4.38\\
$\gamma_{11}$&0.16&0.24&2.36&9.38&1.22&8.31&1.13&1.29&3.64&4.71&3.01&4.52\\
$\gamma_{15}$&0.17&0.23&1.96&9.38&1.08&9.73&0.78&0.93&3.92&4.27&3.32&3.84\\
$\gamma_5$&0.48&0.00&1.22&15.66&0.50&12.90&0.27&0.40&3.69&1.41&4.35&1.59\\
$\gamma_9$&0.51&0.10&1.15&16.35&0.38&12.92&0.37&0.39&16.70&5.62&6.93&2.08\\
$\gamma_2$&0.54&0.07&1.46&20.69&0.58&15.38&0.39&0.40&16.56&5.25&6.91&1.46\\
$\gamma_1$&0.74&0.18&2.15&6.43&1.06&5.97&1.20&0.92&4.10&3.55&4.51&3.90\\
$\gamma_3$&0.91&0.25&1.61&3.03&0.92&3.33&1.57&1.31&2.96&3.66&3.42&4.10\\
$\gamma_4$&1.00&0.01&0.00&6.08&0.00&2.66&0.00&0.00&0.01&0.01&0.17&0.01\\\hline
$\text{C}(\Bgamma)$&1.00&0.99&0.89&0.89&0.95&0.95&0.95&0.95&0.72&0.72&0.74&0.74\\\hline
Eff&$2^{15}$&3276&1906&1906&3212&3212&3276&3276&400&400&416&416\\\hline
Tot&$2^{15}$&3276&3276&3276&6046&6046&3276&3276&3276&3276&3276&3276\\\hline
\end{tabular}
}
\\[1pt]
\caption{\label{rmse.sim.1}Average root mean squared error (RMSE) from the 100 repeated runs of every algorithm on the simulated data (example S.1); the values reported in
the table are RMSE $\times 10^2$ for $p(\gamma_j=1|\By)$. See the caption of Table~\ref{rmse.sim.2} for further details. The corresponding biases are reported in the appendix \ref{sec:further.results} in Table~\ref{bias.sim.3}. The corresponding biases are reported in Table~\ref{bias.sim.1}.}
\end{table}

Table~\ref{rmse.sim.1}, showing the root mean squared errors for different quantities, demonstrate that MJMCMC is outperforming simpler MCMC methods in terms of RM approximations of marginal posterior inclusion probabilities and the total captured mass.  However, the MC approximations seem to be slightly poorer for this example. Whenever both MC and RM approximations are available one should address the  latter since they always have less noise. Comparing MJMCMC results to RM approximations provided by BAS (MC are not available for this method, MJMCMC performed slightly worse when we had 3276 proposals (but 1906 unique models visited). However MJMCMC became equivalent to BAS when we considered 6046 proposals with 3212 unique models visited in MJMCMC (corresponding to similar computational time as BAS). In this example we were not facing a really multiple mode issue having just two modes. All MCMC based methods tended to revisit the same states from time to time and for such a simple example one can hardly ever beat BAS, which never revisits the same solutions and simultaneously draws the models to be estimated in a clever adaptive way with respect to the current marginal posterior inclusion probabilities of individual covariates. 
%

\begin{table}[H]
\resizebox{\textwidth}{!}{%
\begin{tabular}{|c|c|c|c|c|c|Hc|c|c|c|}
\hline
Proposal&Optimizer&Frequency&Type 1&Type 4&Type 3&Type x&Type 5&Type 6&Type 2\\\hline
$q_{g}$&-&$\varrho=0.9836$&0.1176&0.3348&0.2772&0.0000&0.0199&0.2453&0.0042\\\hline
$S$&-&-&$\{2,2\}$&2&$\{2,2\}$&$2$&$1$&$1$&$15$\\\hline
$\rho_j$&-&-&$\widehat{p}(\gamma_j|\By)$&-&-&-&-&-&$\widehat{p}(\gamma_j|\By)$\\\hline
\hline
$q_{l}$&-&0.0164&0&1&0&0&0&0&0\\\hline
$S$&-&-&-&4&$-$&$-$&$-$&$-$&$-$\\\hline
$\rho_i$&-&-&-&-&-&-&-&-&-\\\hline\hline
$q_{o}$&SA&0.5553&0.0788&0.3942&0.1908&0.0000&0.1928&0.1385&0.0040\\\hline
$q_{o}$&GREEDY&0.2404&0.0190&0.3661&0.2111&0.0000&0.2935&0.1046&0.0044\\\hline
$q_{o}$&MTMCMC&0.2043&0.2866&0.1305&0.2329&0.0000&0.1369&0.2087&0.0040\\\hline
$S$&-&-&$\{2,2\}$&2&$\{2,2\}$&$2$&$1$&$1$&$15$\\\hline
$\rho_j$&-&-&$\widehat{p}(\gamma_j|\By)$&-&-&-&-&-&$\widehat{p}(\gamma_j|\By)$\\\hline
\hline
$q_{r}$&-&-&0&0&0&0&0&0&$1$\\\hline
$S$&-&-&-&-&$-$&$-$&$-$&$-$&$15$\\\hline
$\rho_j$&-&-&-&-&-&-&-&-&$0.0010$\\\hline
\end{tabular}
}
\caption{Other tuning parameters of MJMCMC for all proposal types ($q_g, g_l, q_o$, and $q_r$) in example S.1; see Tables~\ref{proposals} and~\ref{proposals1}  for details.} \label{proposalsa1}
\end{table}

\begin{table}[H]
\resizebox{\textwidth}{!}{%
\begin{tabular}{crrrrrrrHrrrr}
\hline
Par&True&TOP&\multicolumn{4}{c}{MJMCMC}&\multicolumn{1}{c}{BAS}&&\multicolumn{2}{c}{$\text{MC}^3$}&\multicolumn{2}{c}{RS}\\\hline
$\Delta$&$\pi_j$&-&RM&MC&RM&MC&RM&unif&MC&RM&MC&RM\\ \hline
$\gamma_{12}$&0.09&-0.29&-2.11&-4.95&-1.19&-5.47&-1.23&-1.35&-0.14&-4.21&0.35&-3.80\\
$\gamma_{14}$&0.10&-0.28&-2.12&-6.58&-1.12&-6.07&-1.14&-1.25&-0.23&-4.23&0.05&-3.89\\
$\gamma_{10}$&0.11&-0.28&-2.30&-6.89&-1.30&-7.64&-1.14&-1.30&-0.10&-4.23&0.11&-4.02\\
$\gamma_8$&0.12&-0.27&-1.96&-6.16&-1.08&-7.69&-0.97&-1.11&0.36&-3.94&-0.51&-3.81\\
$\gamma_6$&0.13&-0.25&-2.24&-8.03&-1.26&-8.33&-1.05&-1.27&-0.65&-4.64&0.06&-4.24\\
$\gamma_7$&0.14&-0.25&-2.05&-7.45&-1.28&-8.37&-1.04&-1.18&-0.13&-4.41&0.08&-4.12\\
$\gamma_{13}$&0.15&-0.24&-2.39&-9.62&-1.35&-8.62&-1.15&-1.24&-0.49&-4.76&0.28&-4.32\\
$\gamma_{11}$&0.16&-0.24&-2.33&-8.69&-1.21&-7.95&-1.13&-1.28&-0.38&-4.59&-0.10&-4.44\\
$\gamma_{15}$&0.17&-0.23&-1.93&-7.64&-1.06&-9.59&-0.78&-0.92&-0.58&-4.15&-0.19&-3.74\\
$\gamma_5$&0.48&0.00&-1.15&-14.18&-0.47&-11.97&-0.25&-0.38&-0.29&-0.94&0.46&-1.17\\
$\gamma_9$&0.51&-0.10&0.78&13.11&0.23&11.96&-0.32&-0.26&-1.79&-2.20&-0.22&-1.53\\
$\gamma_2$&0.54&-0.07&-1.21&-18.43&-0.50&-14.64&0.34&0.27&1.73&0.29&0.35&-0.25\\
$\gamma_1$&0.74&0.18&2.12&4.88&1.04&3.99&1.19&0.91&-0.23&3.39&0.41&3.69\\
$\gamma_3$&0.91&0.25&1.60&-1.79&0.91&0.03&1.56&1.30&-0.40&3.59&-0.14&4.00\\
$\gamma_4$&1.00&0.01&0.00&-5.94&0.00&-2.49&0.00&0.00&0.01&0.01&-0.02&0.01\\\hline
$\text{C}(\Bgamma)$&1.00&0.99&0.89&0.89&0.95&0.95&0.95&0.95&0.72&0.72&0.74&0.74\\\hline
Eff&$2^{15}$&3276&1906&1906&3212&3212&3276&3276&400&400&416&416\\\hline
Tot&$2^{15}$&3276&3276&3276&6046&6046&3276&3276&3276&3276&3276&3276\\\hline
\end{tabular}
}
\\[1pt]
\caption{\label{bias.sim.1}Bias for the 100 simulated runs of every algorithm on the simulated data of experiment S.1; the values reported in
the table are Bias $\times 10^2$ for $p(\gamma_j=1|\By)$. See the caption of Table~\ref{rmse.sim.2} for further details.}
\end{table} 

\end{document}